\documentclass[a4paper,11pt]{article}

\usepackage{amsmath,amssymb}     
\usepackage{color}
\usepackage{graphicx}
\usepackage{subfigure}
\usepackage{cite}                
\usepackage{hyperref}            
\usepackage{multirow,makecell}   

\numberwithin{equation}{section}   

\def \be {\begin{equation}}
\def \ee {\end{equation}}
\def \ba {\begin{array}}
\def \ea {\end{array}}
\def \bea{\begin{eqnarray}}
\def \eea{\end{eqnarray}}
\def \nn {\nonumber}

\def \a {\alpha}
\def \b {\beta}
\def \g {\gamma}

\def \d {\delta}

\def \e {\epsilon}

\def \m {\mu}
\def \n {\nu}
\def \k {\kappa}

\def \L {\Lambda}
\def \s {\sigma}

\def \th {\theta}
\def \vth {\vartheta}

\def \t {\tau}
\def \z {\zeta}

\def \mA {\mathcal A}
\def \mB {\mathcal B}

\def \mN {\mathcal N}

\def \mP {\mathcal P}

\def \p {\partial}
\def \f {\frac}

\def \lt {\left}
\def \rt {\right}

\def \sr {\sqrt}

\def \ph  {\phantom}

\def \hi  {{\hat\imath}}
\def \hj  {{\hat\jmath}}
\def \hk  {{\hat k}}

\def \dd {\mathrm{d}}
\def \ep {\mathrm{e}}
\def \ii {\mathrm{i}}

\def \Tr {{\textrm{Tr}}}

\def \diag {{\textrm{diag}}}

\def \and {{\textrm{and}}}

\def \GY {{\textrm{GY}}}
\def \DT {{\textrm{DT}}}

\begin{document}

\title{\textbf{Construction and classification of novel BPS Wilson loops in quiver Chern-Simons-matter theories}}
\author{
Hao Ouyang\footnote{ouyangh@ihep.ac.cn},
Jun-Bao Wu\footnote{wujb@ihep.ac.cn}~
and
Jia-ju Zhang\footnote{jjzhang@ihep.ac.cn}
}
\date{}

\maketitle

\vspace{-10mm}

\begin{center}
{\it
 Theoretical Physics Division, Institute of High Energy Physics, Chinese Academy of Sciences,\\
19B Yuquan Rd, Beijing 100049, China\\ \vspace{1mm}
Theoretical Physics Center for Science Facilities, Chinese Academy of Sciences,\\19B Yuquan Rd, Beijing 100049, China
}
\vspace{10mm}
\end{center}

\begin{abstract}

  In this paper we construct and classify novel Drukker-Trancanelli (DT) type BPS Wilson loops along infinite straight lines and circles in $\mathcal N=2,3$ quiver superconformal Chern-Simons-matter theories, Aharony-Bergman-Jafferis-Maldacena (ABJM) theory, and $\mathcal N=4$ orbifold ABJM theory. Generally we have four classes of Wilson loops, and all of them preserve the same supersymmetries as the BPS Gaiotto-Yin (GY) type Wilson loops. There are several free complex parameters in the DT type BPS Wilson loops, and for two classes of Wilson loops in ABJM theory and $\mathcal N=4$ orbifold ABJM theory there are supersymmetry enhancements at special values of the parameters. We check that the differences of the DT type and GY type Wilson loops are $Q$-exact with $Q$ being some supercharges preserved by both the DT type and GY type Wilson loops. The results would be useful to calculate vacuum expectation values of the DT type Wilson loops in matrix models if they are still BPS quantum mechanically.

\end{abstract}

\baselineskip 18pt

\thispagestyle{empty}

\newpage

\tableofcontents

\section{Introduction}

BPS (Bogomol'nyi-Prasad-Sommerfield) Wilson loops are important nonlocal objects in supersymmetric gauge theories, and in AdS/CFT correspondence \cite{Maldacena:1997re,Gubser:1998bc,Witten:1998qj} they are dual to probe F-strings/membranes in string/M theory \cite{Maldacena:1998im,Rey:1998ik,Berenstein:1998ij,Drukker:1999zq}, when they are in fundamental representation of the gauge group. BPS Wilson loops in three-dimensional quiver superconformal Chern-Simons-matter (CSM) theories are more complex than those in four-dimensional super Yang-Mills theories. One can use only bosonic fields and construct Gaiotto-Yin (GY) type Wilson loops \cite{Gaiotto:2007qi}, and also one can use both bosonic and fermionic fields and construct Drukker-Trancanelli (DT) type Wilson loops \cite{Drukker:2009hy}.

In $\mN=2$ superconformal CSM theories there are 1/2 BPS GY type Wilson loops along infinite straight lines and circles, and in $\mN=3$ superconformal CSM theories there are 1/3 BPS GY type Wilson loops \cite{Gaiotto:2007qi}.
The Aharony-Bergman-Jafferis-Maldacena (ABJM) theory is an $\mN=6$ superconformal CSM theory with gauge group $U(N)\times U(N)$ and levels $(k,-k)$, and it is dual to M-theory in AdS$_4 \times $S$^7$/Z$_k$ spacetime or type IIA string theory in AdS$_4 \times $CP$^3$ spacetime \cite{Aharony:2008ug}. In ABJM theory, there are 1/6 BPS GY type Wilson loops \cite{Drukker:2008zx,Chen:2008bp,Rey:2008bh} and 1/2 BPS DT type Wilson loops \cite{Drukker:2009hy} along infinite straight lines and circles.
The construction of the latter ones resolved the puzzle about the existence of the half BPS Wilson loop dual to half BPS F-string solution found in \cite{Drukker:2008zx,Rey:2008bh}.
In ABJM theory there are more general BPS DT type Wilson loops along general curves that preserve fewer supersymmetries \cite{Griguolo:2012iq,Cardinali:2012ru,Kim:2013oza,Bianchi:2014laa,Correa:2014aga}.
The $\mN=4$ superconformal CSM theories were constructed in \cite{Gaiotto:2008sd,Hosomichi:2008jd,Imamura:2008dt}, and a special case is the $\mN=4$ orbifold ABJM theory that has gauge group $U(N)^{2r}$ and alternating levels $(k,-k,\cdots,k,-k)$ and is dual to M-theory in AdS$_4 \times $S$^7$/(Z$_r \times $Z$_{rk}$) spacetime \cite{Benna:2008zy,Imamura:2008nn,Terashima:2008ba,Imamura:2008dt}.
Recently, 1/4 BPS GY type Wilson loops and 1/2 BPS DT type BPS Wilson loops in $\mN=4$ orbifold ABJM theory were constructed in \cite{Ouyang:2015qma,Cooke:2015ila}.
BPS Wilson loops in more general $\mN=4$ superconformal CSM theories were also constructed in \cite{Cooke:2015ila}.

As announced in \cite{Ouyang:2015iza}, we found novel DT type BPS Wilson loops in quiver superconformal CSM theories.
In $\mN=2$ quiver superconformal CSM theories, though the supersymmetries are relatively fewer, we found that there still exist
DT type 1/2 BPS Wilson loops. This construction, when applied to theories with more supersymmetries, leads
to DT type Wilson loops preserving two Poincar\'e supercharges (and also two superconformal charges), when the Wilson loops are along straight lines and the parameters in the Wilson loops are not under further constraints.
In ABJM theory and $\mN=4$ CSM theory, supersymmetry (SUSY) enhancements for Wilson loops can appear for special values of these parameters.
We find no SUSY enhancement for Wilson loops in $\mN=3$ CSM theories. This is consistent with the results in the dual M-theory side \cite{Chen:2014gta}.

This paper is an extension of \cite{Ouyang:2015iza}, with calculation details and more examples. We also pay more attention to classification of DT type Wilson loops.
We construct novel DT type BPS Wilson loops along straight lines and circles in several quiver superconformal CSM theories.
We investigate the case of a generic $\mN=2$ quiver superconformal CSM theory with multiple bifundamental and anti-bifundamental matter in section~\ref{s2},
the case of an $\mN=3$ quiver superconformal CSM theory in section~\ref{s4},
the case of ABJM theory in section~\ref{s5},
and the case of $\mN=4$ orbifold ABJM theory in section~\ref{s6}.
We conclude with discussion in section~\ref{s7}.

\section{Generic $\mN$=2 quiver CSM theory}\label{s2}

We consider a generic $\mN=2$ quiver superconformal CSM theory with bifundamental matter. We pick two adjacent nodes in the quiver diagram and the corresponding
gauge groups are $U(N)$ and $U(M)$.
The vector multiplet for gauge group $U(N)$ includes gauge field $A_\mu$, and auxiliary fields $\sigma$, $\chi$, $D$, with $\s$, $D$ being bosonic and $\chi$ being fermionic.
Similarly, for gauge group $U(M)$ we have the vector multiplet with fields $\hat{A}_\mu$, $\hat{\chi}$, $\hat{\sigma}$, $\hat{D}$.
There are multiple matter fields in bifundamental and anti-bifundamental representations in the $\mN=2$ theory.
These multiplets include fields $\phi_i$, $\psi_i$, $F_i$ and $\phi_\hi$, $\psi_\hi$, $F_\hi$, with $i=1,2,\cdots,N_f$, and $\hi=\hat 1,\hat 2,\cdots,\hat N_{\hat f}$. Here $F_i$ and $F_\hi$ are bosonic auxiliary fields.
Note that the number of chiral multiplets in bifundamental representation $N_f$ does not necessarily equal to number of chiral multiplets in anti-bifundamental representation $\hat N_{\hat f}$.
There could be other matter that couple to the two gauge fields, and they will change the on-shell values of $\sigma$ and $\hat{\sigma}$ in the Wilson loops we will construct, but the structure of these Wilson loops will not change.

The SUSY transformations could be found in \cite{Schwarz:2004yj}. For the vector multiplet part, we only need the off-shell SUSY transformations of $A_\mu, \sigma, \hat{A}_\mu, \hat{\sigma}$
\bea \label{offshell}
&& \d A_\m=\f{1}{2} (\bar\chi \g_\m\e+\bar\e \g_\m\chi), ~~~ \d \s=-\f{\ii}{2} (\bar\chi\e+\bar\e\chi),  \nn\\
&& \d \hat A_\m=\f{1}{2} (\bar{\hat\chi} \g_\m\e+\bar\e\g_\m\hat\chi), ~~~ \d \hat\s=-\f{\ii}{2} (\bar{\hat\chi}\e+\bar\e\hat\chi),
\eea
and for the matter part we only need the off-shell transformations
\bea \label{susyne2}
&& \d \phi_i = \ii\bar\e\psi_i, ~~~ \d\bar\phi^i=\ii\bar\psi^i\e, ~~~
   \d \phi_\hi = \ii\bar\e\psi_\hi, ~~~ \d\bar\phi^\hi=\ii\bar\psi^\hi\e \nn\\
&& \d\psi_i = (-\g^\m D_\m \phi_i -\s \phi_i+\phi_i\hat\s)\e - \vth\phi_i  + \ii \bar\e F_i , \nn\\
&& \d\bar\psi^i = \bar\e( \g^\m D_\m \bar\phi^i +\hat\s\bar\phi^i -\bar\phi^i\s) - \bar\vth\bar\phi^i -\ii \e\bar F^i,\\
&& \d\psi_\hi = (-\g^\m D_\m \phi_\hi -\hat\s \phi_\hi+\phi_\hi\s)\e - \vth\phi_\hi + \ii \bar\e F_\hi ,\nn\\
&& \d\bar\psi^\hi = \bar\e( \g^\m D_\m \bar\phi^\hi +\s\bar\phi^\hi -\bar\phi^\hi\hat\s) - \bar\vth\bar\phi^\hi -\ii \e\bar F^\hi. \nn
\eea
The definitions of covariant derivatives are
\bea
&& D_\m \phi_i=\p_\m\phi_i+\ii A_\m\phi_i-\ii\phi_i \hat A_\m, ~~~
   D_\m \bar\phi_i=\p_\m\bar\phi_i+\ii \hat A_\m\bar\phi_i-\ii\bar\phi_i A_\m,     \nn\\
&& D_\m\phi_\hi=\p_\m\phi_\hi+\ii \hat A_\m\phi_\hi-\ii\phi_\hi A_\m, ~~~
   D_\m \bar\phi_\hi=\p_\m\bar\phi_\hi+\ii A_\m\bar\phi_\hi-\ii\bar\phi_\hi \hat A_\m.
\eea
We have SUSY parameters $\e=\th+x^\m\g_\m\vth$, $\bar\e=\bar\th-\bar\vth x^\m\g_\m$, and terms with $\th$, $\bar\th$ denote Poincar\'e SUSY transformations and terms with $\vth$, $\bar\vth$ denote superconformal transformations.

We adopt the spinor convention in \cite{Ouyang:2015ada}. The metric on the three-dimensional Minkowski spacetime is $\eta_{\m\n}=\diag(-++)$, the coordinates are $x^\m=(x^0,x^1,x^2)$, and the gamma matrices are
\be
\g^{\m\phantom{\a}\b}_{\phantom{\m}\a}=(\ii\s^2,\s^1,\s^3),
\ee
with $\s^{1,2,3}$ being Pauli matrices.
Charge conjugate of a spinor is defined as complex conjugate
\be
\bar\th_\a=\th_\a^*.
\ee
The Poincar\'e  supercharges $P$, $\bar P$ and conformal supercharges $S$, $\bar S$ are related to the SUSY transformations as
\be \label{e33}
\d=\ii (\bar \th P+\bar P\th  +  \bar \vth S+\bar S\vth).
\ee
There are constraints
\be\label{e44}
\bar\th=\th^*, ~~~ \bar\vth=\vth^*, ~~~ \bar P=P^*, ~~~ \bar S=S^*.
\ee

After Wick rotation, we get a theory in Euclidean space. In three-dimensional Euclidean space, the metric is $\d_{\m\n}=\diag(+++)$, the coordinates are $x^\m=(x^1,x^2,x^3)$, and the gamma matrices are
\be
\g^{\m\phantom{\a}\b}_{\phantom{\m}\a}=(-\s^2,\s^1,\s^3).
\ee
Spinors $\th$ and $\bar\th$ are generally unrelated. Formally, the SUSY transformations (\ref{offshell}) and (\ref{susyne2}) and the definitions of supercharges (\ref{e33}) still apply to Euclidean version of the $\mN=2$ superconformal CSM theory, but there are no longer the constraints (\ref{e44}).

\subsection{Straight line in Minkowski spacetime}

For BPS Wilson loops along infinite straight lines, the Poincar\'e and conformal supersymmetries are preserved separately, and the discussions of them are similar. So we will only consider Poincar\'e supercharges for Wilson loops along straight lines.

In Minkowski spacetime, one can construct the 1/2 BPS GY type Wilson loop along a timelike infinite straight line $x^\m=\t\d^\m_0$ as \cite{Gaiotto:2007qi}
\bea \label{gy1}
&& W_{\GY}=\mP \exp \lt( -\ii\int\dd\t L_{\GY}(\t) \rt),  \nn\\
&& L_\GY=\lt( \ba{cc} A_\m\dot x^\m+\s |\dot x| & \\ & \hat A_\m\dot x^\m+\hat\s |\dot x| \ea \rt).
\eea
The preserved Poincar\'e supersymmetries can be denoted as
\be \label{susy1}
\g_0\th=\ii\th, ~~~ \bar\th\g_0=\ii\bar\th.
\ee

We construct the DT type Wilson loop along the line $x^\m=\t\d^\m_0$
\bea
&& W_{\DT}=\mP \exp \lt( -\ii\int\dd\t L_{\DT}(\t) \rt),                                                     ~~~
   L_\DT=\lt( \ba{cc} \mA & \bar f_1 \\ f_2 & \hat\mA \ea \rt),                                              \nn\\
&& \mA=A_\m\dot x^\m+\s |\dot x|+ \mB|\dot x|,                                                               ~~~
   \hat\mA=\hat A_\m\dot x^\m+\hat\s |\dot x| + \hat \mB|\dot x|,                                            \nn\\
&& \mB= M^i_{\ph i j}\phi_i\bar\phi^j+M_\hi^{\ph\hi\hj} \bar\phi^\hi\phi_\hj
                                   +M^{ i\hi}\phi_i\phi_\hi  + M_{\hi i}\bar\phi^\hi\bar\phi^i,              \\
&&\hat \mB=  N_i^{\ph ij}\bar\phi^i\phi_j + N^\hi_{\ph\hi\hj}\phi_\hi\bar\phi^\hj
                                                 + N_{i\hi}\bar\phi^i\bar\phi^\hi+N^{\hi i}\phi_\hi\phi_i,   \nn\\
&& \bar f_1=(\bar\zeta^i\psi_i + \bar\psi^\hi \m_\hi ) |\dot x|, ~~~
   f_2= (\bar\psi^i\eta_i+\bar\n^\hi\psi_\hi )|\dot x| .        \nn
\eea
To make it preserve the supersymmetries (\ref{susy1}), it is enough to require that \cite{Lee:2010hk}
\be
\d L_\DT=\p_\t G+\ii[L_\DT,G],
\ee
for some Grassmann odd matrix
\be
G= \lt( \ba{cc} & \bar g_1 \\ g_2 &  \ea \rt).
\ee
Concretely, one needs
\bea \label{lee}
&& \d \mA=\ii(\bar f_1 g_2-\bar g_1 f_2),   \nn\\
&& \d \hat\mA=\ii(f_2 \bar g_1 - g_2 \bar f_1),\\
&& \d \bar f_1 = \p_\t \bar g_1+\ii\mA \bar g_1-\ii\bar g_1\hat\mA,\nn\\
&& \d f_2 = \p_\t g_2+\ii\hat\mA g_2-\ii g_2\mA. \nn
\eea

From variations of $\mA$ and $\hat\mA$ we have
\bea
&& M^i_{\ph ij}=N_j^{\ph ji}, ~~~
   M_\hi^{\ph\hi\hj}=N^\hj_{\ph\hj\hi},~~~
   M^{ i\hi}=N^{\hi i},~~~
   M_{\hi i}=N_{i\hi},  \nn\\
&& ( M^j_{\ph ji}\phi_j + M_{\hi i}\bar \phi^\hi )\th=\eta_i \bar g_1,  ~~~
   ( M_\hj^{\ph\hj\hi}\bar\phi^\hj +M^{i\hi}\phi_i  )\bar\th=-\bar g_1 \bar\n^\hi,\\
&& ( M^i_{\ph ij}\bar\phi^j + M^{i\hi}\phi_\hi)\bar\th=- g_2 \bar\z^i,  ~~~
   ( M_\hi^{\ph\hi\hj}\phi_\hj +M_{\hi i}\bar\phi^i )\th =\m_\hi g_2. \nn
\eea
From variations of $\bar f_1$ and $f_2$ we have
\bea
&& \bar\z^i\g_0=\ii\bar\z^i, ~~~
   \g_0\m_\hi=\ii\m_\hi, ~~~
   \g_0\eta_i=\ii\eta_i, ~~~
   \bar\n^\hi \g_0=\ii\bar\n^\hi,                                    \nn\\
&& \bar g_1=\ii\bar\z^i\th\phi_i-\ii\bar\th\m_\hi\bar\phi^\hi, ~~~
   \mB \bar g_1=\bar g_1 \hat\mB,          \\
&& g_2=-\ii\bar\th\eta_i\bar\phi^i+\ii\bar\n^\hi\th\phi_\hi, ~~~
   \hat \mB g_2=g_2 \mB.                                             \nn
\eea
We have the parameterizations
\bea
&& \bar\zeta^i=\bar\a^i\bar\zeta, ~~~
   \m_\hi=\m\g_\hi, ~~~
   \eta_i=\eta \b_i, ~~~
   \bar\n^\hi=\bar\d^\hi\bar\n,                                     \nn\\
&& \bar\zeta^\a = \bar\n^\a=(1,\ii), ~~~ \eta_\a=\m_\a=(1,-\ii),
\eea
with $\bar\a^i$, $\g_\hi$, $\b_i$ and $\bar\d^\hi$ being complex constants. We have four classes of solutions, and all of them satisfy
\bea
&& M^{i\hi}=M_{\hi i}=\bar\a^i\bar\d^\hi=\g_\hi\b_i=0,  \nn\\
&& M^i_{\ph ij}=2\ii\bar\a^i\b_j, ~~~ M_\hi^{\ph\hi\hj}=2\ii\g_\hi\bar\d^\hj.
\eea

\subsubsection*{Class I}

In the first solution we have
\be
\g_\hi=\bar\d^\hi=0.
\ee
Here $\bar\a^i$ and $\b_j$ are free complex parameters.

We define that
\be
L_\DT-L_\GY=L_B+L_F,
\ee
with $L_B$ being the bosonic part and $L_F$ being the fermionic part.
We want to show that the difference of the DT and GY type BPS Wilson loop is $Q$-exact, i.e.\ that $W_\DT-W_\GY=QV$, with $Q$ being some supercharge that is preserved by both the DT and GY type BPS Wilson loops. For the straight line case, it is enough to show \cite{Drukker:2009hy,Ouyang:2015qma,Cooke:2015ila}
\bea
&& \k \L^2=L_B , ~~~
   Q\L=L_F, ~~~
   Q L_\GY=0,    \nn\\
&& Q L_F=\p_\t(\ii\k\L) +\ii[ L_\GY, \ii\k\L ],
\eea
for some matrix $\L$, factor $\k$ and supercharge $Q$.
Now we have
\bea
&& L_B = 2\ii\bar\a^i\b_j \lt( \ba{cc} \phi_i\bar\phi^j & \\  & \bar\phi^j\phi_i \ea \rt), ~~~
   L_F = \lt( \ba{cc} & \bar\a^i\bar\zeta\psi_i \\ \bar\psi^i\eta\b_i  & \ea \rt), \nn\\
&& \L = \lt( \ba{cc} & \bar\a^i\phi_i \\ \b_i\bar\phi^i & \ea \rt), ~~~
   \k=2\ii, ~~~
   Q=\bar\zeta P+\bar P\eta.
\eea

\subsubsection*{Class II}

In the second solution we have
\be
\bar\a^i=\b_i=0.
\ee
Now we have
\bea
&& L_B =  2\ii\g_\hi\bar\d^\hj\lt( \ba{cc} \bar\phi^\hi\phi_\hj & \\  & \phi_\hj\bar\phi^\hi \ea \rt), ~~~
   L_F = \lt( \ba{cc} & \bar \psi^\hi\m\g_\hi \\ \bar\d^\hi\bar\n\psi_\hi  & \ea \rt), \nn\\
&& \L = \lt( \ba{cc} & \g_\hi\bar\phi^\hi \\ \bar\d^\hi \phi_\hi & \ea \rt), ~~~
   \k=2\ii, ~~~
   Q=\bar\n P+\bar P\m.
\eea

\subsubsection*{Class III}

In the third solution we have
\be
\b_i=\bar\d^\hi=0.
\ee
Now we have
\bea
&& L_B = 0,  ~~~
   L_F = \lt( \ba{cc} & \bar\a^i\bar\zeta\psi_i + \bar \psi^\hi\m\g_\hi \\ 0  & \ea \rt),\\
&& \L = \lt( \ba{cc} & \bar\a^i\phi_i+\g_\hi\bar\phi^\hi  \\  0 & \ea \rt), ~~~
   \k=2\ii, ~~~
   Q=\bar\zeta P+\bar P\m,\nn
\eea

\subsubsection*{Class IV}

In the fourth solution we have
\be
\bar\a^i=\g_\hi=0.
\ee
Now we have
\bea
&& L_B = 0,  ~~~
   L_F = \lt( \ba{cc} & 0 \\ \bar\psi^i\eta\b_i+\bar\d^\hi\bar\n\psi_\hi  & \ea \rt),\\
&& \L = \lt( \ba{cc} & 0 \\ \b_i\bar\phi^i+\bar\d^\hi \phi_\hi & \ea \rt), ~~~
   \k=2\ii, ~~~
   Q=\bar\n P+\bar P\eta.\nn
\eea

\subsection{Circle in Euclidean space}

People are more interested in circular BPS Wilson loops, since they usually have nontrivial vacuum expectation values. It was shown in \cite{Ouyang:2015ada} that there are no spacelike BPS Wilson loops in Minkowski spacetime, and so we have to consider circular BPS Wilson loops in Euclidean version of the $\mN=2$ superconformal CSM theory. For circular BPS Wilson loops there are no separately preserved Poincar\'e and conformal supercharges, and only special combinations of them are preserved. So we have to consider both the Poincar\'e and conformal supercharges for circular Wilson loops.

In Euclidean space we have the 1/2 BPS GY type Wilson loop along the circle $x^\m=(\cos\t,\sin\t,0)$
\bea \label{z1}
&& W_{\GY}=\Tr\mP \exp \lt( -\ii\oint\dd\t L_{\GY}(\t) \rt),\\
&& L_\GY=\lt( \ba{cc} A_\m\dot x^\m-\ii\s |\dot x| & \\ & \hat A_\m\dot x^\m-\ii\hat\s |\dot x| \ea \rt),\nn
\eea
and the preserved supersymmetries are
\be \label{z2}
\vth=\ii \g_3\th, ~~~ \bar\vth=\bar\th\ii\g_3.
\ee

We construct the DT type Wilson loop along $x^\m=(\cos\t,\sin\t,0)$
\bea
&& W_{\DT}=\Tr\mP \exp \lt( -\ii\oint\dd\t L_{\DT}(\t) \rt),                        ~~~
   L_\DT=\lt( \ba{cc} \mA & \bar f_1 \\ f_2 & \hat\mA \ea \rt),                 \nn\\
&& \mA=A_\m\dot x^\m -\ii\s |\dot x|-2( \bar\a^i\b_j \phi_i\bar\phi^j + \g_\hi\bar\d^\hj \bar\phi^\hi\phi_\hj )|\dot x|,  \nn\\
&& \hat\mA=\hat A_\m\dot x^\m -\ii\hat\s |\dot x| - 2( \bar\a^i\b_j \bar\phi^j\phi_i + \g_\hi\bar\d^\hj \phi_\hj\bar\phi^\hi ) |\dot x|,\\
&& \bar f_1=(\bar\a^i\bar\zeta\psi_i + \bar\psi^\hi \m\g_\hi ) |\dot x|, ~~~
   f_2= (\bar\psi^i\eta\b_i+\bar\d^\hi\bar\n\psi_\hi )|\dot x| ,                                                   \nn\\
&& \bar\zeta^\a = \bar\n^\a= (\ep^{\ii\t/2},\ep^{-\ii\t/2}), ~~~ \eta_\a=\m_\a=(\ep^{-\ii\t/2},\ep^{\ii\t/2}).       \nn                                                \nn
\eea
Similar to the case in Minkowski spacetime, we have four classes of solutions that make this circular DT type Wilson loop 1/2 BPS and preserve supersymmetries (\ref{z2}), and all of them satisfy
\be
\bar\a^i\bar\d^\hi=\g_\hi\b_i=0.
\ee

\subsubsection*{Class I}

In the first solution we have
\be
\g_\hi=\bar\d^\hi=0.
\ee
To show that the difference of the DT and GY type BPS Wilson loop is $Q$-exact, for the circle case we have to show \cite{Drukker:2009hy,Ouyang:2015qma,Cooke:2015ila}
\bea
&& \k \L^2=L_B , ~~~
   \k(0)=\k(2\pi), ~~~
   \L(0)=-\L(2\pi),\\
&& Q\L=L_F, ~~~
   Q L_\GY=0, ~~~
   Q L_F=\p_\t(\ii\k\L) +\ii[ L_\GY, \ii\k\L ],\nn
\eea
for some matrix $\L$, factor $\k$ and supercharge $Q$.
Now we have
\bea
&& L_B = -2\bar\a^i\b_j \lt( \ba{cc} \phi_i\bar\phi^j & \\  & \bar\phi^j\phi_i \ea \rt), ~~~
   L_F = \lt( \ba{cc} & \bar\a^i\bar\zeta\psi_i \\ \bar\psi^i\eta\b_i  & \ea \rt),               \nn\\
&& \L = \ep^{\ii\t/2}\lt( \ba{cc} & \bar\a^i\phi_i \\ \b_i\bar\phi^i & \ea \rt), ~~~
   \k=-2\ep^{-\ii\t},                                                                            \\
&& Q=\bar a (P+\ii\g_3 S)+(\bar P + \bar S \ii \g_3) b, ~~~
   \bar a^\a=(1,0), ~~~
   b_\a=(0,1).                                                                                   \nn
\eea

\subsubsection*{Class II}

In the second solution we have
\be
\bar\a^i=\b_i=0.
\ee
Now we have
\bea
&& L_B =  -2\g_\hi\bar\d^\hj\lt( \ba{cc} \bar\phi^\hi\phi_\hj & \\  & \phi_\hj\bar\phi^\hi \ea \rt), ~~~
   L_F = \lt( \ba{cc} & \bar \psi^\hi\m\g_\hi \\ \bar\d^\hi\bar\n\psi_\hi  & \ea \rt),                          \nn\\
&& \L = \ep^{\ii\t/2}\lt( \ba{cc} & \g_\hi\bar\phi^\hi \\ \bar\d^\hi \phi_\hi & \ea \rt), ~~~
   \k=-2\ep^{-\ii\t},                                                                                                     \\
&& Q=\bar a (P+\ii\g_3 S)+(\bar P + \bar S \ii \g_3) b, ~~~
   \bar a^\a=(1,0), ~~~
   b_\a=(0,1).                                                                                                  \nn
\eea

\subsubsection*{Class III}

In the third solution we have
\be
\b_i=\bar\d^\hi=0.
\ee
Now we have
\bea
&& L_B = 0,  ~~~
   L_F = \lt( \ba{cc} & \bar\a^i\bar\zeta\psi_i + \bar \psi^\hi\m\g_\hi \\ 0  & \ea \rt),         \nn\\
&& \L = \ep^{\ii\t/2}\lt( \ba{cc} & \bar\a^i\phi_i+\g_\hi\bar\phi^\hi  \\  0 & \ea \rt), ~~~
   \k=-2\ep^{-\ii\t},                                                                             \\
&& Q=\bar a (P+\ii\g_3 S)+(\bar P + \bar S \ii \g_3) b, ~~~
   \bar a^\a=(1,0), ~~~
   b_\a=(0,1).                                                                                    \nn
\eea

\subsubsection*{Class IV}

In the fourth solution we have
\be
\bar\a^i=\g_\hi=0.
\ee
Now we have
\bea
&& L_B = 0,  ~~~
   L_F = \lt( \ba{cc} & 0 \\ \bar\psi^i\eta\b_i+\bar\d^\hi\bar\n\psi_\hi  & \ea \rt),             \nn\\
&& \L = \ep^{\ii\t/2}\lt( \ba{cc} & 0 \\ \b_i\bar\phi^i+\bar\d^\hi \phi_\hi & \ea \rt), ~~~
   \k=-2\ep^{-\ii\t},                                                                             \\
&& Q=\bar a (P+\ii\g_3 S)+(\bar P + \bar S \ii \g_3) b, ~~~
   \bar a^\a=(1,0), ~~~
   b_\a=(0,1).                                                                                    \nn
\eea

\section{$\mN=3$ quiver CSM theory}\label{s4}

We consider an $\mN=3$ quiver superconformal CSM theory and pick two adjacent nodes in the quiver diagram. The first vector multiplet includes gauge field $A_\m$ and auxiliary fields $\s^a_{~b}$, $\chi^a_{~b}$, $\xi$, with $a,b,\cdots=1,2$ being indices of the $SU(2)$ R-symmetry. Here $\s^a_{~b}$ is bosonic and $\chi^a_{~b}$, $\xi$ are fermionic.
We have the constraints
\bea
&& \s^a_{~a}=0, ~~~ \s^a_{~b}=(\s^b_{~a})^\dagger,         \nn\\
&& \chi^a_{~a}=0, ~~~ \chi^a_{~b}=(\s^b_{~a})^\dagger,     \\
&& \xi=\xi^\dagger,                                      \nn
\eea
with $\dagger$ being not only hermitian conjugate of color index but also complex conjugate of spinor index.
Similarly, for the second vector multiplet we have fields $\hat A_\m$, $\hat\s^a_{~b}$, $\hat\chi^a_{~b}$, $\hat\xi$.
There are bifundamental matter fields $\phi^{ia}$, $\psi^{ia}$ with $i,j,\cdots=1,2,\cdots,N_f$ denoting indices of flavor.
From the results in \cite{Kao:1993gs,Gaiotto:2007qi} we rewrite the off-shell SUSY transformations with manifest $SU(2)$ R-symmetry
\bea \label{e36}
&& \d A_\m=\f{\ii}{2} \chi^a_{~b}\g_\m\e^b_{~a}, ~~~
   \d \hat A_\m=\f{\ii}{2} \hat\chi^a_{~b}\g_\m\e^b_{~a},                                        \nn\\
&& \d \s^a_{~b}=-\f{1}{2} ( \chi^a_{~c} \e^c_{~b} -\e^a_{~c}\chi^c_{~b} ) +\ii\xi \e^a_{~b},   \nn\\
&& \d \hat \s^a_{~b}=-\f{1}{2} ( \hat\chi^a_{~c} \e^c_{~b} -\e^a_{~c}\hat\chi^c_{~b} ) +\ii\hat\xi \e^a_{~b},   \\
&& \d\phi^{ia}=\ii\e^a_{~b}\psi^{ib}, ~~~
   \d\bar\phi_{ia}=\ii\bar\psi_{ib}\e^b_{~a}, \nn\\
&& \d\psi^{ia}=-\g^\m\e^a_{~b}D_\m\phi^{ib} -\vth^a_{~b}\phi^{ib} -\e^b_{~c} ( \s^a_{~b}\phi^{ic}-\phi^{ic}\hat\s^a_{~b} ),   \nn\\
&& \d\bar\psi_{ia}=\e^b_{~a}\g^\m D_\m\bar\phi_{ib} -\vth^b_{~a}\bar\phi_{ib} -\e^c_{~b}(\bar\phi_{ic}\s^b_{~a}-\hat\s^b_{~a}\bar\phi_{ic}). \nn
\eea
We have covariant derivatives
\bea
&& D_\m \phi^{ia}=\p_\m\phi^{ia}+\ii A_\m\phi^{ia}-\ii\phi^{ia} \hat A_\m,                  \nn\\
&& D_\m \bar\phi_{ia}=\p_\m\bar\phi_{ia}+\ii \hat A_\m\bar\phi_{ia}-\ii\bar\phi_{ia} A_\m.
\eea
We have the SUSY transformation parameter $\e^a_{~b}=\th^a_{~b}+x^\m\g_\m\vth^a_{~b}$, and  there are constraints
\be
\th^a_{~a}=0, ~~~ \th^a_{~b}=(\th^b_{~a})^*, ~~~
\vth^a_{~a}=0, ~~~ \vth^a_{~b}=(\vth^b_{~a})^*.
\ee
Note that each of $\th^a_{~b}$ and $\vth^a_{~b}$ has the degrees of freedom of one Dirac spinor and one Majorana spinor.
The Poincar\'e and conformal supercharges are defined as
\be  \label{e37}
\d =\ii(\th^a_{~b}P^b_{~a}+\vth^a_{~b}S^b_{~a})=\ii (P^a_{~b} \th^b_{~a} + S^a_{~b} \vth^b_{~a}),
\ee
with constraints
\be
P^a_{~a}=0, ~~~ P^a_{~b}=(P^b_{~a})^*, ~~~ S^a_{~a}=0, ~~~ S^a_{~b}=(S^b_{~a})^*.
\ee

In Euclidean space, the SUSY transformations (\ref{e36}) and definitions of supercharges (\ref{e37}) still apply, but the constraints for the SUSY parameters and supercharges are
\be
\th^a_{~a}=\vth^a_{~a}=0, ~~~ P^a_{~a}=S^a_{~a}=0.
\ee

\subsection{Straight line in Minkowski spacetime}

We consider BPS GY and DT type Wilson loops along timelike infinite straight lines in Minkowski spacetime.

\subsubsection{GY type Wilson loop}

We construct a GY type Wilson loop along the infinite straight line $x^\m=\t\d^\m_0$ as
\bea \label{ne3gy}
&& W_{\GY}=\mP \exp \lt( -\ii\int\dd\t L_{\GY}(\t) \rt),               ~~~
   L_\GY=\lt( \ba{cc} \mA_\GY & \\ & \hat \mA_\GY \ea \rt),            \\
&& \mA_{GY}=A_\m \dot x^\m + R^a_{~b}\s^b_{~a}|\dot x|,                ~~~
   \hat\mA_{GY}=\hat A_\m \dot x^\m + S^a_{~b}\hat\s^b_{~a}|\dot x|.   \nn
\eea
Without loss of generality we set $R^a_{~a}=S^a_{~a}=0$. We take SUSY transformation $\d \mA_\GY=0$ and get
\be
\g_0\th^a_{~b}=\ii ( R^a_{~c}\th^c_{~b} - \th^a_{~c}R^c_{~b}), ~~~
R^a_{~b}\th^b_{~a}=0,
\ee
whose complex conjugates are
\be
\g_0\th^a_{~b}=\ii ( R^{\dagger a}_{~~c}\th^c_{~b} - \th^a_{~c}R^{\dagger c}_{~~b}), ~~~
R^{\dagger a}_{~~b}\th^b_{~a}=0.
\ee
Here the matrix $R^\dagger$ is the hermitian conjugate of $R$, i.e.\ $R^{\dagger a}_{~~b}=(R^b_{~a})^*$.
We define
\be
R=B+\ii C, ~~~ B=\f{R+R^\dagger}{2}, ~~~ C=-\ii\f{R-R^\dagger}{2},
\ee
with $B$ and $C$ being traceless hermitian matrices, i.e.\ $B^a_{~a}=0$, $B=B^\dagger$, $C^a_{~a}=0$, $C=C^\dagger$. Then we get
\be
\g_0\th^a_{~b}=\ii ( B^{a}_{~c}\th^c_{~b} - \th^a_{~c}B^{c}_{~b}), ~~~
C^{a}_{~c}\th^c_{~b}=\th^{a}_{~c}C^c_{~b}, ~~~
B^a_{~b}\th^b_{~a}=C^a_{~b}\th^b_{~a}=0.
\ee
We make an $SU(2)$ R-symmetry transformation and set $B^a_{~b}=\diag(b,-b)$ with $b$ being real. We have
\be
\g_0\th^1_{~2}=2\ii b\th^1_{~2}, ~~~
\g_0\th^2_{~1}=-2\ii b\th^2_{~1}, ~~~
\th^1_{~1}=\th^2_{~2}=0.
\ee
Since the eigenvalues of $\g_0$ must be $\pm\ii$, for $\th^1_{~2}\neq0$ we have $b=\pm\f{1}{2}$, and without loss of generality we choose $b=\f12$. Then we get $C^a_{~b}=0$. From SUSY transformation $\d \hat\mA_\GY=0$, we have
\be
\g_0\th^a_{~b}=\ii ( S^a_{~c}\th^c_{~b} - \th^a_{~c}S^c_{~b}), ~~~
R^a_{~b}\th^b_{~a}=0.
\ee
Then we get
\be
(R^a_{~c}-S^a_{~c})\th^c_{~b} = \th^a_{~c}( R^c_{~b} - S^c_{~b}),
\ee
and this leads to $R^a_{~b}=S^a_{~b}$.

In summary we have the GY type 1/3 BPS Wilson loop (\ref{ne3gy}) with $R^a_{~b}=S^a_{~b}=\diag(\f12,-\f12)$, and the preserved supersymmetries are
\be\label{susyne4}
\g_0\th^1_{~2}=\ii \th^1_{~2}, ~~~
\g_0\th^2_{~1}=-\ii \th^2_{~1}, ~~~
\th^1_{~1}=\th^2_{~2}=0.
\ee
This is just the Wilson loop that was constructed in \cite{Gaiotto:2007qi}. Here we show that it is the only kind of BPS GY type Wilson loops along timelike infinite straight lines, up to some R-symmetry transformations.

\subsubsection{DT type Wilson loop}

Along $x^\m=\t\d^\m_0$ we construct the DT type Wilson loop
\bea \label{ne3dt}
&& W_{\DT}=\mP \exp \lt( -\ii\int\dd\t L_{\DT}(\t) \rt),                   ~~~
   L_\DT=\lt( \ba{cc} \mA & \bar f_1 \\ f_2 & \hat\mA \ea \rt),            \nn\\
&& \mA=\mA_{\GY} + M_{ia}^{~~jb}\phi^{ia}\bar\phi_{jb} |\dot x|,           ~~~
   \bar f_1=\bar\z_{ia}\psi^{ia}|\dot x|,             \\
&& \hat\mA=\hat\mA_{\GY} + N^{ia}_{~~jb}\bar\phi_{ia}\phi^{jb}|\dot x|,     ~~~
   f_2=\bar\psi_{ia}\eta^{ia}|\dot x|.                                      \nn
\eea
We want the Wilson loop to preserve the supersymmetries (\ref{susyne4}).
From variations of $\mA$ and $\hat\mA$ we get
\be
M_{ia}^{~~jb}=N^{jb}_{~~ia}, ~~~
\eta^{ia} \bar g_1=M_{jb}^{~~ic}\th^a_{~c}\phi^{jb}, ~~~
g_2 \bar\z_{ia}=-M_{ic}^{~~jb}\th^c_{~a}\bar\phi_{jb}.
\ee
From variations of $\bar f_1$ and $f_2$ we have
\bea
&& \bar\z_{i1}\g_0=\ii\bar\z_{i1}, ~~~ \bar\z_{i2}\g_0=-\ii\bar\z_{i2}, ~~~
   \g_0\eta^{i1}=\ii \eta^{i1}, ~~~ \g_0\eta^{i2}=-\ii \eta^{i2},         \nn\\
&& \bar g_1=\bar\z_{ia}\g_0\th^a_{~b}\phi^{ib}, ~~~
   M_{ia}^{~~jb}\phi^{ia}\bar\phi_{jb} \bar g_1=\bar g_1 M^{~~ia}_{jb}\bar\phi_{ia}\phi^{jb},                                       \\
&& g_2=-\bar\phi_{ib}\th^b_{~a}\g_0\eta^{ia}, ~~~
   M^{~~ia}_{jb}\bar\phi_{ia}\phi^{jb}g_2 = g_2 M_{ia}^{~~jb}\phi^{ia}\bar\phi_{jb}. \nn
\eea
Then we get the parameterizations
\bea
&& \bar\z_{i1}=\bar\a_i \bar\z, ~~~ \bar \z^\a=(1,\ii),    ~~~
   \bar\z_{i2}=\bar\g_i \bar\m, ~~~ \bar \m^\a=(-\ii,-1),   \nn\\
&& \eta^{i1}=\eta\b^i, ~~~ \eta_\a=(1,-\ii),               ~~~
   \eta^{i2}=\n\d^i, ~~~ \n_\a=(-\ii,1),
\eea
as well as
\bea
&& \eta^{ia}\bar\z_{jc}\g_0\th^c_{~b}=M_{jb}^{~~ic}\th^a_{~c},                      ~~~
   M_{ia}^{~~jb}\bar\z_{kd}\g_0\th^d_{~c}=M_{kc}^{~~jb}\bar\z_{id}\g_0\th^d_{~a},                      \\
&& \th^b_{~c}\g_0\eta^{jc}\bar\z_{ia}=M_{ic}^{~~jb}\th^c_{~a},   ~~~
   M_{jb}^{~~ia}\th^c_{~d}\g_0\eta^{kd}=M_{jb}^{~~kc}\th^a_{~d}\g_0\eta^{id}.       \nn
\eea
We have four classes of solutions, and all of them must satisfy
\bea
&& M_{i1}^{~~j2}=M_{i2}^{~~j1}=\bar\a_i\d^j=\bar\g_i\b^j=0, \nn\\
&& M_{i1}^{~~j1}=2\ii\bar\g_i\d^j, ~~~
   M_{i2}^{~~j2}=2\ii\bar\a_i\b^j.
\eea

\subsection*{Class I}

In the first solution we have
\be
\bar\g_i=\d^i=0.
\ee
Here $\bar\a_i$ and $\b^i$ are free complex parameters.
Now we have
\bea
&& L_B =  2\ii\bar\a_i\b^j \lt( \ba{cc} \phi^{i2}\bar\phi_{j2} & \\  & \bar\phi_{j2}\phi^{i2} \ea \rt), ~~~
   L_F = \lt( \ba{cc} & \bar\a_i\bar\z\psi^{i1} \\ \bar\psi_{i1}\eta\b^i  & \ea \rt), \nn\\
&& \L = \lt( \ba{cc} & \bar\a_i\phi^{i2}  \\ \b^i\bar\phi_{i2} & \ea \rt), ~~~
   \k=2\ii, ~~~
   Q=\bar\zeta P^1_{~2} + P^2_{~1}\eta,
\eea
and this makes that $W_\DT-W_\GY=QV$.

\subsection*{Class II}

In the second solution we have
\be
\bar\a_i=\b^i=0.
\ee
Now we have
\bea
&& L_B =  2\ii\bar\g_i\d^j \lt( \ba{cc} \phi^{i1}\bar\phi_{j1} & \\  & \bar\phi_{j1}\phi^{i1} \ea \rt), ~~~
   L_F = \lt( \ba{cc} & \bar\g_i\bar\m\psi^{i2} \\\bar\psi_{i2}\n\d^i  & \ea \rt), \nn\\
&& \L = \lt( \ba{cc} & \bar\g_i\phi^{i1} \\ \d^i\bar\phi_{i1} & \ea \rt), ~~~
   \k=2\ii, ~~~
   Q=\bar\m P^2_{~1} + P^1_{~2}\n.
\eea

\subsection*{Class III}

In the third solution we have
\be
\b^i=\d^i=0.
\ee
Now we have
\bea
&& L_B = 0, ~~~
   L_F = \lt( \ba{cc} & \bar\a_i\bar\z\psi^{i1}+\bar\g_i\bar\m\psi^{i2} \\ 0 & \ea \rt), \nn\\
&& \L = \lt( \ba{cc} & \bar\a_i\phi^{i2} + \bar\g_i\phi^{i1} \\ 0 & \ea \rt), ~~~
   \k=-2\ii, ~~~
   Q=\bar\zeta P^1_{~2} + \bar\m P^2_{~1}.
\eea

\subsection*{Class IV}

In the fourth solution we have
\be
\bar\a_i=\bar\g_i=0.
\ee
Now we have
\bea
&& L_B =  0, ~~~
   L_F = \lt( \ba{cc} & 0 \\ \bar\psi_{i1}\eta\b^i + \bar\psi_{i2}\n\d^i  & \ea \rt), \nn\\
&& \L = \lt( \ba{cc} & 0 \\ \b^i\bar\phi_{i2} + \d^i\bar\phi_{i1} & \ea \rt), ~~~
   \k=-2\ii, ~~~
   Q=P^2_{~1}\eta + P^1_{~2}\n.
\eea

\subsection{Circle in Euclidean space}

We consider circular BPS GY and DT type Wilson loops in Euclidean space.

\subsubsection{GY type Wilson loop}

We get the 1/3 BPS GY type Wilson loop along the circle $x^\m=(\cos\t,\sin\t,0)$
\bea
&& W_{\GY}=\Tr \mP \exp \lt( -\ii\oint\dd\t L_{\GY}(\t) \rt),               ~~~
   L_\GY=\lt( \ba{cc} \mA_\GY & \\ & \hat \mA_\GY \ea \rt),\\
&& \mA_{GY}=A_\m \dot x^\m + R^a_{~b}\s^b_{~a}|\dot x|,                ~~~
   \hat\mA_{GY}=\hat A_\m \dot x^\m + R^a_{~b}\hat\s^b_{~a}|\dot x|,\nn
\eea
with $R^a_{~b}=\diag(-\ii/2,\ii/2)$, and the preserved supersymmetries are
\be \label{e38}
\vth^1_{~2}=\ii\g_3\th^1_{~2}, ~~~ \vth^2_{~1}=-\ii\g_3\th^2_{~1}, ~~~ \th^1_{~1}=\th^2_{~2}=\vth^1_{~1}=\vth^2_{~2}=0.
\ee

\subsubsection{DT type Wilson loop}

We construct the DT type Wilson loop along $x^\m=(\cos\t,\sin\t,0)$
\bea
&& W_{\DT}=\Tr\mP \exp \lt( -\ii\oint\dd\t L_{\DT}(\t) \rt), ~~~
   L_\DT=\lt( \ba{cc} \mA & \bar f_1 \\ f_2 & \hat\mA \ea \rt),                                                \nn\\
&& \mA=\mA_{\GY}-2(\bar\g_i\d^j \phi^{i1}\bar\phi_{j1}+\bar\a_i\b^j \phi^{i2}\bar\phi_{j2}) |\dot x|,          \nn\\
&& \hat\mA=\hat\mA_{\GY} -2(\bar\g_i\d^j\bar\phi_{j1}\phi^{i1}+\bar\a_i\b^j\bar\phi_{j2}\phi^{i2})|\dot x|,    \\
&& \bar f_1=(\bar\a_i\bar\z\psi^{1a}+\bar\g_i\bar\m\psi^{2a})|\dot x|, ~~~
      \bar\z^\a=(\ep^{\ii\t/2},\ep^{-\ii\t/2}), ~~~
   \bar\m^\a=(\ep^{\ii\t/2},-\ep^{-\ii\t/2}),                                                                      \nn\\
&& f_2=(\bar\psi_{i1}\eta\b^{i} + \bar\psi_{i2}\n\d^{i} )|\dot x|, ~~~
   \eta_\a=(\ep^{-\ii\t/2},\ep^{\ii\t/2}),  ~~~
   \n_\a=(-\ep^{-\ii\t/2},\ep^{\ii\t/2}).                                                                           \nn
\eea
We want the Wilson loop to preserve the supersymmetries (\ref{e38}).
We have four classes of solutions, and all of them must satisfy
\be
\bar\a_i\d^j=\bar\g_i\b^j=0.
\ee

\subsection*{Class I}

In the first solution we have
\be
\bar\g_i=\d^i=0.
\ee
Here $\bar\a_i$ and $\b^i$ are free complex parameters.
Now we have
\bea
&& L_B =  -2\bar\a_i\b^j \lt( \ba{cc} \phi^{i2}\bar\phi_{j2} & \\  & \bar\phi_{j2}\phi^{i2} \ea \rt), ~~~
   L_F = \lt( \ba{cc} & \bar\a_i\bar\z\psi^{i1} \\ \bar\psi_{i1}\eta\b^i  & \ea \rt),                      \nn\\
&& \L = \ep^{\ii\t/2}\lt( \ba{cc} & \bar\a_i\phi^{i2}  \\ \b^i\bar\phi_{i2} & \ea \rt), ~~~
   \k=-2\ep^{-\ii\t},                                                                                      \\
&& Q=\bar a (P^1_{~2}+\ii\g_3S^1_{~2}) + (P^2_{~1}+S^2_{~1}\ii\g_3)b, ~~~ \bar a^\a=(1,0), ~~~ b_\a=(0,1), \nn
\eea
and this makes that $W_\DT-W_\GY=QV$.

\subsection*{Class II}

In the second solution we have
\be
\bar\a_i=\b^i=0.
\ee
Now we have
\bea
&& L_B =  -2\bar\g_i\d^j \lt( \ba{cc} \phi^{i1}\bar\phi_{j1} & \\  & \bar\phi_{j1}\phi^{i1} \ea \rt), ~~~
   L_F = \lt( \ba{cc} & \bar\g_i\bar\m\psi^{i2} \\\bar\psi_{i2}\n\d^i  & \ea \rt),                            \nn\\
&& \L = \ep^{\ii\t/2}\lt( \ba{cc} & \bar\g_i\phi^{i1} \\ \d^i\bar\phi_{i1} & \ea \rt), ~~~
   \k=-2\ep^{-\ii\t},                                                                                                   \\
&& Q=\bar a (P^2_{~1}-\ii\g_3S^2_{~1}) + (P^1_{~2}-S^1_{~2}\ii\g_3)b, ~~~ \bar a^\a=(1,0), ~~~ b_\a=(0,1).    \nn
\eea

\subsection*{Class III}

In the third solution we have
\be
\b^i=\d^i=0.
\ee
Now we have
\bea
&& L_B = 0, ~~~
   L_F = \lt( \ba{cc} & \bar\a_i\bar\z\psi^{i1}+\bar\g_i\bar\m\psi^{i2} \\ 0 & \ea \rt),                   \nn\\
&& \L = \ep^{\ii\t/2}\lt( \ba{cc} & \bar\a_i\phi^{i2} + \bar\g_i\phi^{i1} \\ 0 & \ea \rt), ~~~
   \k=2\ep^{-\ii\t},                                                                                               \\
&& Q=\bar a (P^1_{~2}+\ii\g_3S^1_{~2} + P^2_{~1}-\ii\g_3 S^2_{~1}), ~~~ \bar a^\a=(1,0). \nn
\eea

\subsection*{Class IV}

In the fourth solution we have
\be
\bar\a_i=\bar\g_i=0.
\ee
Now we have
\bea
&& L_B =  0, ~~~
   L_F = \lt( \ba{cc} & 0 \\ \bar\psi_{i1}\eta\b^i + \bar\psi_{i2}\n\d^i  & \ea \rt),                      \nn\\
&& \L = \ep^{\ii\t/2} \lt( \ba{cc} & 0 \\ \b^i\bar\phi_{i2} + \d^i\bar\phi_{i1} & \ea \rt), ~~~
   \k=2\ep^{-\ii\t},                                                                                               \\
&& Q=(P^1_{~2}-S^1_{~2}\ii\g_3 + P^2_{~1}+S^2_{~1}\ii\g_3)b, ~~~ b_\a=(0,1). \nn
\eea

\section{ABJM theory}\label{s5}

The ABJM theory \cite{Aharony:2008ug} is an $\mN=6$ superconformal CSM theory, with gauge group $U(N)\times U(N)$ and levels $(k,-k)$. There are matter fields $\phi_I$, $\psi^I$, with $I=1,2,3,4$ being the index of $SU(4)$ R-symmetry.
The SUSY transformations of ABJM theory are \cite{Gaiotto:2008cg,Hosomichi:2008jb,Terashima:2008sy,Bandres:2008ry}
\bea\label{susytransfabjm}
&& \d A_\m=\f{2\pi}{k} \lt( \phi_I\bar\psi_J\g_\m\e^{IJ} +\bar\e_{IJ}\g_\m\psi^J\bar\phi^I \rt), \nn\\
&& \d\hat A_\m=\f{2\pi}{k} \lt( \bar\psi_J\g_\m\phi_I\e^{IJ}+\bar\e_{IJ}\bar\phi^I\g_\m\psi^J \rt), \nn\\
&& \d\phi_I=\ii\bar\e_{IJ}\psi^J, ~~~ \d\bar\phi^I=\ii\bar\psi_J\e^{IJ},\\
&& \d\psi^I=\g^\m\e^{IJ}D_\m\phi_J + \vth^{IJ}\phi_J
            -\f{2\pi}{k}\e^{IJ} \lt( \phi_J\bar\phi^K\phi_K-\phi_K\bar\phi^K\phi_J \rt)
            -\f{4\pi}{k}\e^{KL}\phi_K\bar\phi^I\phi_L, \nn\\
&& \d\bar\psi_I=-\bar\e_{IJ}\g^\m D_\m\bar\phi^J + \bar\vth_{IJ}\bar\phi^J
                +\f{2\pi}{k}\bar\e_{IJ} \lt( \bar\phi^J\phi_K\bar\phi^K-\bar\phi^K\phi_K\bar\phi^J \rt)
                +\f{4\pi}{k}\bar\e_{KL}\bar\phi^K\phi_I\bar\phi^L. \nn
\eea
The definitions of covariant derivatives are
\bea
&& D_\m\phi_J =\p_\m \phi_J +\ii A_\m \phi_J -\ii \phi_J \hat A_\m ,\nn\\
&& D_\m\bar\phi^J=\p_\m\bar\phi^J +\ii \hat A_\m \bar\phi^J -\ii \bar\phi^J  A_\m.
\eea
We have SUSY parameters $\e^{IJ}=\th^{IJ}+x^\m\g_\m\vth^{IJ}$ and $\bar\e_{IJ}=\bar\th_{IJ}-\bar\vth_{IJ}x^\m\g_\m$, with constraints
\bea
&& \th^{IJ}=-\th^{JI}, ~~~ (\th^{IJ})^*=\bar \th_{IJ}, ~~~ \bar\th_{IJ}=\f{1}{2}\e_{IJKL}\th^{KL},    \nn\\
&& \vth^{IJ}=-\vth^{JI}, ~~~ (\vth^{IJ})^*=\bar \vth_{IJ}, ~~~ \bar\vth_{IJ}=\f{1}{2}\e_{IJKL}\vth^{KL}.
\eea
Symbol $\e_{IJKL}$ is totally antisymmetric with $\e_{1234}=1$.
The supercharges are defined as
\be \label{e39}
\d=\f{\ii}{2} (\bar \th_{IJ} P^{IJ} + \bar \vth_{IJ} S^{IJ})
  =\f{\ii}{2} (\bar P_{IJ} \th^{IJ} + \bar S_{IJ} \vth^{IJ}),
\ee
with the constraints
\bea
&& P^{IJ}=-P^{JI}, ~~~ (P^{IJ})^*=\bar P_{IJ}, ~~~ \bar P_{IJ}=\f{1}{2}\e_{IJKL}P^{KL},       \nn\\
&& S^{IJ}=-S^{JI}, ~~~ (S^{IJ})^*=\bar S_{IJ}, ~~~ \bar S_{IJ}=\f{1}{2}\e_{IJKL}S^{KL}.
\eea

In Euclidean space, the SUSY transformations (\ref{susytransfabjm}) and definitions of supercharges (\ref{e39}) still apply, but the constraints for the SUSY parameters and supercharges are
\bea
&& \th^{IJ}=-\th^{JI}, ~~~ \bar\th_{IJ}=\f{1}{2}\e_{IJKL}\th^{KL},       ~~~
   \vth^{IJ}=-\vth^{JI}, ~~~ \bar\vth_{IJ}=\f{1}{2}\e_{IJKL}\vth^{KL},  \nn\\
&& P^{IJ}=-P^{JI}, ~~~ \bar P_{IJ}=\f{1}{2}\e_{IJKL}P^{KL},             ~~~
   S^{IJ}=-S^{JI}, ~~~ \bar S_{IJ}=\f{1}{2}\e_{IJKL}S^{KL}.
\eea

\subsection{Straight line in Minkowski spacetime}

We consider BPS GY and DT type Wilson loops along timelike infinite straight lines in Minkowski spacetime.

\subsubsection{GY type Wilson loop}

A general GY type Wilson loop along the line $x^\m=\t\d^\m_0$ takes the form
\bea\label{abjmgy}
&& W_\GY=\mP \exp \lt( -\ii\int\dd\t L_\GY(\t) \rt), ~~~
   L_\GY=\lt( \ba{cc} \mA_\GY & \\ & \hat\mA_\GY \ea \rt),                                   \nn\\
&& \mA_\GY=A_\m \dot x^\m +\f{2\pi}{k} R^I_{\ph{I}J} \phi_I\bar\phi^J |\dot x|,           ~~~
   \hat\mA_\GY=\hat A_\m \dot x^\m +\f{2\pi}{k} S_I^{\ph{I}J} \bar\phi^I\phi_J |\dot x|.
\eea
The SUSY transformation $\d\mA_\GY=0$ leads to
\be
\g_0\th^{IJ}=-\ii R^I_{\ph IK}\th^{KJ}, ~~~ \bar\th_{IJ}\g_0=-\ii R^K_{\ph KI}\bar\th_{KJ}.
\ee
Taking complex conjugate of the second equation we get
\be
\g_0\th^{IJ}=-\ii R^{\dagger I}_{\ph{\dagger I}K}\th^{KJ},
\ee
with the matrix $R^\dagger$ being the hermitian conjugate of $R$, i.e.\ $R^{\dagger I}_{\ph{\dagger I}J}=(R^{J}_{\ph{J}I})^*$. We define
\be
R=B+\ii C, ~~~ B=\f{R+R^\dagger}{2}, ~~~ C=-\ii\f{R-R^\dagger}{2},
\ee
with $B$ and $C$ being hermitian matrices, i.e.\ $B=B^\dagger$ and $C=C^\dagger$.
Then we have
\be
\g_0\th^{IJ}=-\ii B^I_{\ph IK}\th^{KJ}, ~~~ C^I_{\ph IK}\th^{KJ}=0.
\ee
We use $SU(4)$ transformation of the R-symmetry to make the hermitian matrix $B$ diagonal
\be
B^I_{\ph IJ}=\diag(b_1,b_2,b_3,b_4),
\ee
with $b_1$, $b_2$, $b_3$, and $b_4$ being real. Then we get
\be
\g_0 \th^{IJ}=-\ii b_I \th^{IJ},
\ee
with no summation of index $I$ in the right-hand side. Note that the eigenvalues of the matrix $\g_0$ can only be $\pm\ii$. Without loss of generality, we suppose
\be
\g_0 \th^{12}=\ii\th^{12}, ~~~ \th^{12} \neq 0,
\ee
and then using $\th^{12*}=\th^{34}$ we have
\be
\g_0 \th^{34}=-\ii\th^{34}, ~~~ \th^{34} \neq 0.
\ee
We get $b_1=b_2=-1$, $b_3=b_4=1$, and then we have
\be
\th^{13}=\th^{14}=\th^{23}=\th^{24}=0.
\ee
Then we get $C^I_{\ph IJ}=0$, i.e.\ that $R^I_{\ph IJ}$ is a hermitian matrix.
The SUSY transformation $\d\hat\mA_\GY=0$ leads to
\be
\g_0\th^{IJ}=-\ii S_K^{\ph KI}\th^{KJ}, ~~~ \bar\th_{IJ}\g_0=-\ii S_I^{\ph IK}\bar\th_{KJ},
\ee
from which we get
\be
(R^I_{\ph IK}-S_K^{\ph KI})\th^{KJ}=0, ~~~ (R^K_{\ph KI}-S_I^{\ph IK})\bar\th_{KJ}=0.
\ee
We get $S_I^{\ph IJ}=R^J_{\ph JI}$.

In summary, we have the GY type 1/6 BPS Wilson loop (\ref{abjmgy}) with $R^I_{\ph IJ}=S_J^{\ph JI}=\diag(-1,-1,1,1)$, and the preserved supersymmetries are
\bea\label{susy2}
&& \g_0\th^{12}=\ii\th^{12}, ~~~ \g_0\th^{34}=-\ii\th^{34},  \nn\\
&& \th^{13}=\th^{14}=\th^{23}=\th^{24}=0.
\eea
This is just the Wilson loop that was constructed in \cite{Drukker:2008zx,Chen:2008bp,Rey:2008bh}. Here we show that this is the only form of GY type BPS Wilson loop along a timelike straight line, up to some $SU(4)$ R-symmetry transformation. Especially, we do not require that $R^I_{\ph I J}$ or $S_I^{\ph I J}$ is a hermitian matrix \emph{a priori}, and we show that it is the result of supersymmetric invariance.

\subsubsection{DT type Wilson loop}

We want a DT type Wilson loop that preserves the same supersymmetries (\ref{susy2}). A general DT type Wilson loop is \cite{Drukker:2009hy}
\bea
&& W_\DT = \mP \exp \lt( -\ii\int\dd\t L_\DT(\t) \rt), ~~~
   L_\DT = \lt( \ba{cc} \mA & \bar f_1 \\ f_2 & \hat\mA \ea \rt),                                   \nn\\
&& \mA = \mA_\GY +\f{2\pi}{k} M^I_{\ph{I}J} \phi_I\bar\phi^J |\dot x|, ~~~
   \bar f_1=\sr{\f{2\pi}{k}}\bar\zeta_I\psi^I |\dot x|,          \\
&& \hat\mA = \hat\mA_\GY +\f{2\pi}{k} N_I^{\ph{I}J} \bar\phi^I\phi_J |\dot x|, ~~~
   f_2=\sr{\f{2\pi}{k}}\bar\psi_I\eta^I |\dot x|.                                   \nn
\eea
From variations of $\mA$ and $\hat\mA$ we have
\be \label{e2}
M^I_{\ph IJ}=N_J^{\ph JI}, ~~~
\sr{\f{2\pi}{k}}M^J_{\ph JK}\phi_J\th^{KI}=\eta^I \bar g_1, ~~~
\sr{\f{2\pi}{k}}M^J_{\ph JK}\bar\phi^K\bar\th_{JI}= -g_2\bar\zeta_I.
\ee
From variations of $\bar f_1$ and $f_2$ we have
\bea \label{fermiabjm}
&& \bar\z_{1,2}\g_0=\ii\bar\z_{1,2}, ~~~ \bar\z_{3,4}\g_0=-\ii\bar\z_{3,4}, ~~~
   \g_0 \eta_{1,2}=\ii \eta_{1,2}, ~~~ \g_0 \eta_{3,4}=-\ii \eta_{3,4},                    \nn\\
&& \bar g_1=-\sr{\f{2\pi}{k}}\bar\z_I\g_0\th^{IJ}\phi_J, ~~~
   M^I_{\ph IJ}\bar\z_L\gamma_0\th^{LK}= M^K_{\ph KJ} \bar\z_L\gamma_0\theta^{LI},                                  \\
&& g_2=\sr{\f{2\pi}{k}}\bar\th_{IJ}\g_0\eta^I\bar\phi^J,         ~~~
   M^I_{\ph IJ} \bar\th_{KL}\gamma_0\eta^L = M^I_{\ph IK} \bar\th_{JL}\gamma_0\eta^L.      \nn
   \eea
We make the parameterizations
\bea
&& \bar\z_{1,2}=\bar\a_{1,2}\bar\z, ~~~ \bar \z^\a=(1,\ii),    ~~~
   \bar\z_{3,4}=\bar\g_{3,4}\bar\m, ~~~ \bar \m^\a=(-\ii,-1),   \nn\\
&& \eta^{1,2}=\eta\b^{1,2}, ~~~ \eta_\a=(1,-\ii),               ~~~
   \eta^{3,4}=\n\d^{3,4}, ~~~ \n_\a=(-\ii,1).
\eea
Then equations (\ref{e2}) become
\be
M^I_{\ph IK} \th^{KJ} = - \eta^J \bar\z_K\g_0\th^{KI}, ~~~
M^K_{\ph KI} \bar\th_{KJ} = -\bar\th_{KI}\g_0\eta^K \bar\zeta_J.
\ee
We find four classes of solutions, and all of them satisfy
\bea
&&  M^I_{\ph IJ}=2\ii\lt(\ba{cccc}
   \bar\a_2\b^2  & -\bar\a_2\b^1 & &\\
   -\bar\a_1\b^2 & \bar\a_1\b^1  & & \\
     &  & \bar\g_4\d^4  & -\bar\g_4\d^3 \\
     &  & -\bar\g_3\d^4 & \bar\g_3\d^3
   \ea\rt),                                    \\
&& \bar\a_{1,2}\d^{3,4}=\bar\g_{3,4}\b^{1,2}=0. \nn
\eea

\subsubsection*{Class I}

In the first solution we have
\be \label{e40}
\bar\g_{3,4}=\d^{3,4}=0.
\ee
Note that there are four free complex parameters $\bar\a_{1,2}$ and $\b^{1,2}$.

Now we have
\bea
&& L_B =  \f{2\pi}{k} M^I_{\ph IJ}\lt( \ba{cc}\phi_I\bar\phi^I & \\ & \bar\phi^J\phi_I \ea \rt), ~~~
   L_F = \sr{\f{2\pi}{k}} \lt( \ba{cc} & \bar\zeta_I\psi^I \\ \bar\psi_I\eta^I  & \ea \rt), \nn\\
&& \L = \sr{\f{2\pi}{k}} \lt( \ba{cc} & \bar\a_2 \phi_1-\bar\a_1 \phi_2 \\ \b^2 \bar\phi^1-\b^1 \bar\phi^2 & \ea \rt), ~~~
   \k=2\ii, ~~~
   Q=\bar\zeta P^{12}+\bar P_{12}\eta.
\eea
This shows that the difference between the DT type Wilson loop and GY type Wilson loop is $Q$-exact.

We wonder if there can be more preserved supersymmetries than (\ref{susy2}) for the DT BPS Wilson loop, at least for some special values of the parameters $\bar\a_{1,2}$ and $\b^{1,2}$. We solve the equations (\ref{lee}) of the Wilson loop with general SUSY transformation parameters $\th^{IJ}$ and $\bar\th_{IJ}$. From variations of $\bar f_1$ and $f_2$ we get
\bea
&& \g_0 \bar\a_I \th^{IJ}=\ii \bar\a_I \th^{IJ}, ~~~
   \bar g_1=-\ii \sr{\f{2\pi}{k}}\bar\z\th^{IJ}\bar\a_I\phi_J,       \nn\\
&& \b^I \bar\th_{IJ}\g_0=\ii \b^I \bar\th_{IJ}, ~~~
   g_2=\ii \sr{\f{2\pi}{k}}\bar\th_{IJ}\eta\b^I\bar\phi^J.
\eea
Then from variations of $\mA$ and $\hat\mA$ we get
\be \label{e3}
\g_0\th^{IJ}+\ii U^I_{\ph IK}\th^{KJ}=2\bar\a_K\b^J\th^{KI}, ~~~
\bar\th_{IJ}\g_0+\ii U^K_{\ph KI}\bar\th_{KJ}=2\bar\a_J\b^K\bar\th_{KI}.
\ee
We define $\a^I=\bar\a_I^*$ and $\bar\b_I=\b^{I*}$. Using the fact that $\th^{12}=\th^{34*}$, $\th^{13}=-\th^{24*}$ and $\th^{23}=\th^{14*}$, we can show that equations (\ref{e3}) are equivalent to
\bea
&& \bar\a_1\b^1+\bar\a_2\b^2=-\ii, ~~~
   \g_0\th^{12}=\ii\th^{12}, \nn\\
&& \g_0\th^{13}=-(\ii+2\bar\a_1\b^1)\th^{13}-2\bar\a_2\b^1\th^{23},  \nn\\
&& \g_0\th^{23}=-2\bar\a_1\b^2\th^{13}+(\ii+2\bar\a_1\b^1)\th^{23},\\
&& (\bar\a_1\b^1+\bar\b_1\a^1)\th^{13}+(\bar\a_2\b^1+\bar\b_2\a^1)\th^{23}=0,\nn\\
&& (\bar\a_1\b^2+\bar\b_1\a^2)\th^{13}-(\bar\a_1\b^1+\bar\b_1\a^1)\th^{23}=0.\nn
\eea
Note that when an equation is satisfied, for consistency so is its complex conjugate. When there is SUSY enhancement for the Wilson loop, there must be nonvanishing solution for at least one of $\th^{13}$ and $\th^{23}$, and this requires that
\be
\lt| \begin{array}{cc}
       \bar\a_1\b^1+\bar\b_1\a^1   &  \bar\a_2\b^1+\bar\b_2\a^1 \\
       \bar\a_1\b^2+\bar\b_1\a^2  &  -(\bar\a_1\b^1+\bar\b_1\a^1)
     \end{array}
  \rt|=-(\bar\a_1\b^1+\bar\b_1\a^1)^2-|\bar\a_1\b^2+\bar\b_1\a^2|^2=0.
\ee
This leads to
\be
\bar\a_1\b^1+\bar\b_1\a^1=\bar\a_1\b^2+\bar\b_1\a^2=0.
\ee
Then we get the solution
\be \label{e4}
\b^I=-\f{\ii}{\bar\a_J \a^J} \a^I, ~~~ \bar\a_I\neq0.
\ee
It can be checked that supersymmetries are enhanced to 1/2 BPS, and the preserved supersymmetries are
\be \label{ss1}
\g_0 \bar\a_I \th^{IJ}=\ii\bar\a_I\th^{IJ}, ~~~
\g_0 \e_{IJKL}\a^J \th^{KL}=-\ii\e_{IJKL}\a^J \th^{KL}.
\ee
Note that the two equations are consistent, since they are complex conjugates of each other.
In this case we have
\be
U^I_{\ph IJ}=\d^I_J-2\ii \bar\a_J\b^I=\d^I_J-\f{2\bar\a_J\a^I}{\bar\a_K\a^K}.
\ee
The Wilson loop is global $SU(3)$ R-symmetry invariant. When $\bar \a_2=0$, it is just Wilson loop that was constructed in \cite{Drukker:2009hy}.

So for general parameters the Wilson loop with (\ref{e40}) is 1/6 BPS, and the preserved supersymmetries are (\ref{susy2}). When there is also (\ref{e4}), the Wilson loop is enhanced to 1/2 BPS, and the preserved supersymmetries are (\ref{ss1}).

\subsubsection*{Class II}

In the second solution the calculation is parallel to that of the first one. We have
\be \label{e41}
\bar\a_{1,2}=\b^{1,2}=0,
\ee
and the Wilson loop is 1/6 BPS and preserves the supersymmetries (\ref{susy2}).
Now we have
\bea
&& L_B =  \f{2\pi}{k} M^I_{\ph IJ}\lt( \ba{cc}\phi_I\bar\phi^J & \\  & \bar\phi^J\phi_I \ea \rt), ~~~
   L_F = \sr{\f{2\pi}{k}} \lt( \ba{cc} & \bar\zeta_I\psi^I \\ \bar\psi_I\eta^I  & \ea \rt), \nn\\
&& \L = \sr{\f{2\pi}{k}} \lt( \ba{cc} & \bar\g_4 \phi_3-\bar\g_3 \phi_4 \\ \d^4 \bar\phi^3-\d^3 \bar\phi^4 & \ea \rt), ~~~
   \k=2\ii, ~~~
   Q=\bar\m P^{34}+\bar P_{34}\n.
\eea

We want to find if there is SUSY enhancement for the Wilson loop with (\ref{e41}). From variations of $\bar f_1$ and $f_2$ we get
\bea
&& \g_0 \bar \g_I \th^{IJ}=-\ii\bar\g_I\th^{IJ}, ~~~
   \bar g_1=\ii\sr{\f{2\pi}{k}}\bar\m\th^{IJ}\bar\g_I\phi_J,  \nn\\
&& \d^I\bar\th_{IJ}\g_0=-\ii\d^I\bar\th_{IJ}, ~~~
   g_2=-\ii\sr{\f{2\pi}{k}}\bar\th_{IJ}\n\d^I\bar\phi^J.
\eea
From variations of $\mA$ and $\hat\mA$ we get
\be \label{e5}
\g_0\th^{IJ}+\ii U^I_{\ph IK}\th^{KJ}=2\bar\g_K\d^J\th^{KI}, ~~~
\bar\th_{IJ}\g_0+\ii U^K_{\ph KI}\bar\th_{KJ}=2\bar\g_J\d^K\bar\th_{KI}.
\ee
We define $\g^I=\bar\g_I^*$ and $\bar\d_I=\d^{I*}$. We can show that equations (\ref{e5}) are equivalent to
\bea
&& \bar\g_3\d^3+\bar\g_4\d^4=\ii, ~~~
   \g_0\th^{12}=\ii\th^{12}, \nn\\
&& \g_0\th^{13}=(\ii-2\bar\g_3\d^3)\th^{13}-2\bar\g_4\d^3\th^{14},  \nn\\
&& \g_0\th^{14}=-2\bar\g_3\d^4\th^{13}-(\ii-2\bar\g_3\d^3)\th^{14},\\
&& (\bar\g_3\d^3+\bar\d_3\g^3)\th^{13}+(\bar\g_4\d^3+\bar\d_4\g^3)\th^{14}=0,\nn\\
&& (\bar\g_3\d^4+\bar\d_3\g^4)\th^{13}-(\bar\g_3\d^3+\bar\d_3\g^3)\th^{14}=0.\nn
\eea
For the existence of nonvanishing solution for at least one of $\th^{13}$ and $\th^{14}$, we have
\be
\lt|  \begin{array}{cc}
     \bar\g_3\d^3+\bar\d_3\g^3  &  \bar\g_4\d^3+\bar\d_4\g^3 \\
     \bar\g_3\d^4+\bar\d_3\g^4  &  -(\bar\g_3\d^3+\bar\d_3\g^3)
      \end{array}
 \rt|=-(\bar\g_3\d^3+\bar\d_3\g^3)^2-|\bar\g_3\d^4+\bar\d_3\g^4|^2=0.
\ee
This leads to
\be
\bar\g_3\d^3+\bar\d_3\g^3=\bar\g_3\d^4+\bar\d_3\g^4=0.
\ee
Then we get the solution
\be
\d^I=\f{\ii}{\bar\g_J \g^J} \g^I, ~~~ \bar\g_I\neq0.
\ee
It can be checked that supersymmetries are enhanced to 1/2 BPS, and the preserved supersymmetries are\footnote{Please distinguish the gamma matrix $\g_0$ from the complex parameters $\bar\g_I$, $\g^I$.}
\be
\g_0 \bar\g_I \th^{IJ}=-\ii\bar\g_I\th^{IJ}, ~~~
\g_0 \e_{IJKL}\g^J \th^{KL}=\ii\e_{IJKL}\g^J \th^{KL}.
\ee
At present we have
\be
U^I_{\ph IJ}=-\d^I_J-2\ii\d^I\bar\g_J=-\d^I_J+\f{2\g^I\bar\g_J}{\bar\g_K\g^K}.
\ee
The Wilson loop is global $SU(3)$ R-symmetry invariant.

\subsubsection*{Class III}

In the third solution we have
\be \label{e42}
\b^{1,2}=\d^{3,4}=0.
\ee
The Wilson loop is 1/6 BPS Wilson loop and preserves the supersymmetries (\ref{susy2}).
Now we have
\bea
&& L_B = 0, ~~~
   L_F = \sr{\f{2\pi}{k}} \lt( \ba{cc} & \bar\zeta_I\psi^I \\ 0  & \ea \rt), \nn\\
&& \L = \sr{\f{2\pi}{k}} \lt( \ba{cc} & \bar\a_2 \phi_1-\bar\a_1 \phi_2+\bar\g_4 \phi_3-\bar\g_3 \phi_4 \\ 0 & \ea \rt), \nn\\
&&   \k=-2\ii, ~~~
   Q=\bar\zeta P^{12}+\bar\m P^{34}.
\eea

For the Wilson loop with (\ref{e42}), we have $f_2=g_2=0$ and so $\d\mA=\d\hat\mA=0$. This leads to
\be
\g_0 \th^{IJ}=-\ii R^I_{\ph IK}\phi^{KJ}, ~~~ \bar\th_{IJ}\g_0=-\ii R^K_{\ph KI} \bar\phi_{KJ}.
\ee
The solution is (\ref{susy2}). So there is no SUSY enhancement for this solution.

\subsubsection*{Class IV}

The fourth solution is like the third one. We have
\be \label{e43}
\bar\a_{1,2}=\bar\g_{3,4}=0,
\ee
and the 1/6 BPS Wilson loop that preserves the supersymmetries (\ref{susy2}).
Now we have
\bea
&& L_B =  0, ~~~
   L_F = \sr{\f{2\pi}{k}} \lt( \ba{cc} & 0 \\ \bar\psi_I\eta^I  & \ea \rt), \nn\\
&& \L = \sr{\f{2\pi}{k}} \lt( \ba{cc} & 0 \\ \b^2 \bar\phi^1-\b^1 \bar\phi^2+\d^4 \bar\phi^3-\d^3 \bar\phi^4 & \ea \rt),\\
&& \k=-2\ii, ~~~
   Q=\bar P_{12}\eta +\bar P_{34}\n.\nn
\eea

For the Wilson loop with (\ref{e43}), we have $\bar f_1=\bar g_1=0$ and so $\d\mA=\d\hat\mA=0$. This leads to the supersymmetries (\ref{susy2}). There is no SUSY enhancement for this solution.

In \cite{Cooke:2015ila}, exactly $1/3$ BPS Wilson loops in ABJM theory were  anticipated to exist. However we do not find $1/3$ BPS  DT type Wilson loops within the Wilson loops at least preserving supersymmetries (\ref{susy2}).

\subsection{Circle in Euclidean space}

We consider circular BPS GY and DT type Wilson loops in Euclidean space.

\subsubsection{GY type Wilson loop}

In Euclidean space, one can construct the circular 1/6 BPS GY type Wilson loop along $x^\m=(\cos\t,\sin\t,0)$
\bea
&& W_\GY=\Tr \mP\exp \lt( -\ii\oint\dd\t L_\GY(\t) \rt), ~~~
   L_\GY=\lt( \ba{cc} \mA_\GY & \\ & \hat\mA_\GY \ea \rt),                                 \nn\\
&& \mA_\GY=A_\m \dot x^\m +\f{2\pi}{k} R^I_{\ph{I}J} \phi_I\bar\phi^J |\dot x|,           ~~~
   \hat\mA_\GY=\hat A_\m \dot x^\m +\f{2\pi}{k} R^I_{\ph{I}J}\bar\phi^J\phi_I |\dot x|.
\eea
with $R^I_{\ph{I}J}=\diag(\ii,\ii,-\ii,-\ii)$.
The preserved supersymmetries are
\bea \label{susyabjmcircle}
&& \vth^{12}=\ii\g_3\th^{12}, ~~~ \vth^{34}=-\ii\g_3\th^{34},   \nn\\
&& \th^{13}=\th^{14}=\th^{23}=\th^{24}=0,                       \\
&& \vth^{13}=\vth^{14}=\vth^{23}=\vth^{24}=0.                   \nn
\eea

\subsubsection{DT type Wilson loop}

We also construct the DT type Wilson loop along $x^\m=(\cos\t,\sin\t,0)$
\bea
&& W_\DT =\Tr \mP \exp \lt( -\ii\oint\dd\t L_\DT(\t) \rt), ~~~
   L_\DT = \lt( \ba{cc} \mA & \bar f_1 \\ f_2 & \hat\mA \ea \rt),                         \nn\\
&& \mA = A_\m\dot x^\m +\f{2\pi}{k} U^I_{\ph{I}J} \phi_I\bar\phi^J |\dot x|, ~~~
   \bar f_1=\sr{\f{2\pi}{k}}( \bar\a_I\bar\zeta + \bar\g_I\bar\m )\psi^I |\dot x|,         \nn\\
&& \hat\mA = \hat A_\m\dot x^\m +\f{2\pi}{k} U^I_{\ph{I}J}\bar\phi^J\phi_I |\dot x|, ~~~
   f_2=\sr{\f{2\pi}{k}}\bar\psi_I(\eta\b^I+\n\d^I) |\dot x|,                               \\
&& U^I_{\ph IJ}=\lt(\ba{cccc}
\ii-2\bar\a_2\b^2  & 2\bar\a_2\b^1      & &\\
2\bar\a_1\b^2      & \ii-2\bar\a_1\b^1  & & \\
                   &                    & -\ii-2\bar\g_4\d^4  & 2\bar\g_4\d^3 \\
                   &                    & 2\bar\g_3\d^4       & -\ii-2\bar\g_3\d^3
\ea\rt),                                                                                    \nn\\
&& \bar\a_I=(\bar\a_1,\bar\a_2,0,0), ~~~
   \bar\z^\a=(\ep^{\ii\t/2},\ep^{-\ii\t/2}), ~~~
   \b^I=(\b^1,\b^2,0,0), ~~~
   \eta_\a=(\ep^{-\ii\t/2},\ep^{\ii\t/2}),                                                   \nn\\
&& \bar\g_I=(0,0,\bar\g_3,\bar\g_4), ~~~
   \bar\m^\a=(\ep^{\ii\t/2},-\ep^{-\ii\t/2}), ~~~
   \d^I=(0,0,\d^3,\d^4), ~~~
   \n_\a=(-\ep^{-\ii\t/2},\ep^{\ii\t/2}). \nn
\eea
Similar to the case in Minkowski spacetime, we have four classes of solutions that make this DT type Wilson loop 1/6 BPS, and the preserved supersymmetries are (\ref{susyabjmcircle}). All of them must satisfy
\be
\bar\a_I\d^J=\bar\g_I\b^J=0.
\ee

\subsubsection*{Class I}

In the first solution we have
\be
\bar\g_I=\d^I=0.
\ee
Now we have
\bea
&& L_B =  \f{2\pi}{k} M^I_{\ph IJ}\lt( \ba{cc}\phi_I\bar\phi^I & \\ & \bar\phi^J\phi_I \ea \rt), ~~~
   L_F = \sr{\f{2\pi}{k}} \lt( \ba{cc} & \bar\a_I\bar\zeta\psi^I \\ \bar\psi_I\eta\b^I  & \ea \rt),                                               \nn\\
&& \L = \sr{\f{2\pi}{k}} \ep^{\ii\t/2}\lt( \ba{cc} & \bar\a_2 \phi_1-\bar\a_1 \phi_2 \\ \b^2 \bar\phi^1-\b^1 \bar\phi^2 & \ea \rt), ~~~
   \k=-2\ep^{-\ii\t},                                                                                                                     \\
&& Q=\bar a (P^{12} + \ii\g_3S^{12}) + (\bar P_{12}+\bar S_{12}\ii\g_3)b, ~~~
   \bar a^\a=(1,0), ~~~ b_\a=(0,1). \nn
\eea
This shows that the difference between the DT type Wilson loop and GY type Wilson loop is $Q$-exact.
When
\be
\b^I=\f{\ii}{\bar\a_J \a^J} \a^I, ~~~ \bar\a_I\neq0,
\ee
the Wilson loop becomes 1/2 BPS, and the preserved supersymmetries are
\be
\bar\a_I \vth^{IJ}=\ii\g_3 \bar\a_I\th^{IJ},               ~~~
\e_{IJKL}\a^J \vth^{KL}=-\ii\g_3\e_{IJKL}\a^J \th^{KL}.
\ee

\subsubsection*{Class II}

In the second solution we have
\be
\bar\a_I=\b^I=0.
\ee
Now we have
\bea
&& L_B =  \f{2\pi}{k} M^I_{\ph IJ}\lt( \ba{cc}\phi_I\bar\phi^J & \\  & \bar\phi^J\phi_I \ea \rt), ~~~
   L_F = \sr{\f{2\pi}{k}} \lt( \ba{cc} & \bar\g_I\bar\m\psi^I \\ \bar\psi_I\n\d^I  & \ea \rt), \nn\\
&& \L = \sr{\f{2\pi}{k}} \ep^{\ii\t/2}\lt( \ba{cc} & \bar\g_4 \phi_3-\bar\g_3 \phi_4 \\ \d^4 \bar\phi^3-\d^3 \bar\phi^4 & \ea \rt), ~~~
   \k=-2\ep^{-\ii\t},                                                                                                                     \\
&& Q=\bar a (P^{34} - \ii\g_3S^{34}) + (\bar P_{34}-\bar S_{34}\ii\g_3)b, ~~~
   \bar a^\a=(1,0), ~~~ b_\a=(0,1). \nn
\eea
When
\be
\d^I=-\f{\ii}{\bar\g_J \g^J} \g^I, ~~~ \bar\g_I\neq0,
\ee
the Wilson loop is 1/2 BPS, and the preserved supersymmetries are
\be
\bar\g_I \vth^{IJ}=-\ii\g_3 \bar\g_I\th^{IJ},            ~~~
\e_{IJKL}\g^J \vth^{KL}=\ii\g_3 \e_{IJKL}\g^J \th^{KL}.
\ee

\subsubsection*{Class III}

In the third solution we have
\be
\b^I=\d^I=0.
\ee
Now we have
\bea
&& L_B = 0, ~~~
   L_F = \sr{\f{2\pi}{k}} \lt( \ba{cc} & (\bar\a_I\bar\zeta + \bar\g_I\bar\m)\psi^I \\ 0  & \ea \rt),  ~~~
   \k=2\ep^{-\ii\t},     \nn\\
&& \L = \sr{\f{2\pi}{k}} \ep^{\ii\t/2}\lt( \ba{cc} & \bar\a_2 \phi_1-\bar\a_1 \phi_2+\bar\g_4 \phi_3-\bar\g_3 \phi_4 \\ 0 & \ea \rt),                                                                                                                \\
&& Q=\bar a (P^{12} + \ii\g_3S^{12} + P^{34} - \ii\g_3S^{34}) , ~~~
   \bar a^\a=(1,0). \nn
\eea
There is no SUSY enhancement for this solution.

\subsubsection*{Class IV}

In the fourth solution we have
\be
\bar\a_I=\bar\g_I=0.
\ee
Now we have
\bea
&& L_B = 0, ~~~
   L_F = \sr{\f{2\pi}{k}} \lt( \ba{cc} & 0 \\ \bar\psi_I(\eta\b^I + \n\d^I ) & \ea \rt), ~~~
   \k=2\ep^{-\ii\t},   \nn\\
&& \L = \sr{\f{2\pi}{k}} \ep^{\ii\t/2} \lt( \ba{cc} & 0 \\ \b^2 \bar\phi^1-\b^1 \bar\phi^2+\d^4 \bar\phi^3-\d^3 \bar\phi^4 & \ea \rt),                                                                                                                     \\
&& Q=(\bar P_{12}+\bar S_{12}\ii\g_3+\bar P_{34}-\bar S_{34}\ii\g_3)b, ~~~
   b_\a=(0,1). \nn
\eea
There is no SUSY enhancement for this solution.

\section{$\mN$=4 orbifold ABJM theory}\label{s6}

The calculation for the Wilson loops in $\mN$=4 orbifold ABJM theory is very similar to the case of ABJM theory. It will be brief in this section.

Orbifolding the ABJM theory with gauge group $U(rN)\times U(rN)$ and levels $(k,-k)$ by $Z_r$,
one gets the $\mN=4$ superconformal CSM theory with gauge group $U(N)^{2r}$ and Chern-Simons levels $(k, -k, \cdots, k, -k)$ \cite{Benna:2008zy}.
We get the SUSY transformations of the $\mN=4$ orbifold ABJM theory from those of ABJM theory
\bea \label{susytransfne4}
&& \d\phi_i^{(2\ell+1)} = \ii\bar\e_{i\hi}\psi^\hi_{(2\ell+1)}, ~~~
   \d\phi_\hi^{(2\ell)} = -\ii\bar\e_{i\hi}\psi^i_{(2\ell)}, ~~~
   \d\bar\phi^i_{(2\ell+1)} = \ii\bar\psi_\hi^{(2\ell+1)}\e^{i\hi}, ~~~
   \d\bar\phi^\hi_{(2\ell)} = -\ii\bar\psi_i^{(2\ell)}\e^{i\hi},  \nn\\
&& \d A_\m^{(2\ell+1)}=\f{2\pi}{k} \lt[ \lt( \phi_i^{(2\ell+1)}\bar\psi_\hi^{(2\ell+1)}
                                            -\phi_\hi^{(2\ell)}\bar\psi_i^{(2\ell)} \rt) \g_\m \e^{i\hi}
                                        +\bar\e_{i\hi}\g_\m \lt( \psi^\hi_{(2\ell+1)}\bar\phi^i_{(2\ell+1)}
                                                         -\psi^i_{(2\ell)}\bar\phi^\hi_{(2\ell)} \rt)\rt],    \nn\\
&& \d\hat A_\m^{(2\ell)}=\f{2\pi}{k} \lt[ \lt( \bar\psi_\hi^{(2\ell-1)}\phi_i^{(2\ell-1)}
                                              -\bar\psi_i^{(2\ell)}\phi_\hi^{(2\ell)} \rt) \g_\m\e^{i\hi}
+\bar\e_{i\hi}\g_\m \lt( \bar\phi^i_{(2\ell-1)}\psi^\hi_{(2\ell-1)}-\bar\phi^\hi_{(2\ell)}\psi^i_{(2\ell)} \rt) \rt], \nn\\
&& \d\psi^i_{(2\ell)}=\g^\m\e^{i\hi}D_\m\phi_\hi^{(2\ell)} + \vth^{i\hi}\phi_\hi^{(2\ell)}
            -\f{2\pi}{k}\e^{i\hi} \lt(  \phi_\hi^{(2\ell)}\bar\phi^j_{(2\ell-1)}\phi_j^{(2\ell-1)} \rt.   \nn\\
&& \phantom{\d\psi^i_{(2\ell)}=}   \lt. +\phi_\hi^{(2\ell)}\bar\phi^\hj_{(2\ell)}\phi_\hj^{(2\ell)}
                                        -\phi_j^{(2\ell+1)}\bar\phi^j_{(2\ell+1)}\phi_\hi^{(2\ell)}
                                        -\phi_\hj^{(2\ell)}\bar\phi^\hj_{(2\ell)}\phi_\hi^{(2\ell)}  \rt)   \nn\\
&& \phantom{\d\psi^i_{(2\ell)}=}
            -\f{4\pi}{k}\e^{j\hj} \lt(  \phi_j^{(2\ell+1)}\bar\phi^i_{(2\ell+1)}\phi_\hj^{(2\ell)}
                                        -\phi_\hj^{(2\ell)}\bar\phi^i_{(2\ell-1)}\phi_j^{(2\ell-1)} \rt),   \nn\\
&& \d\psi^\hi_{(2\ell+1)}= -\g^\m\e^{i\hi}D_\m\phi_i^{(2\ell+1)} - \vth^{i\hi}\phi_i^{(2\ell+1)}
+\f{2\pi}{k}\e^{i\hi} \lt( \phi_i^{(2\ell+1)}\bar\phi^j_{(2\ell+1)}\phi_j^{(2\ell+1)} \rt.                \nn\\
&& \phantom{\d\psi^\hi_{(2\ell+1)}=} \lt.  +\phi_i^{(2\ell+1)}\bar\phi^\hj_{(2\ell+2)}\phi_\hj^{(2\ell+2)}
                                           -\phi_j^{(2\ell+1)}\bar\phi^j_{(2\ell+1)}\phi_i^{(2\ell+1)}
                                           -\phi_\hj^{(2\ell)}\bar\phi^\hj_{(2\ell)}\phi_i^{(2\ell+1)} \rt) \nn\\
&& \phantom{\d\psi^\hi_{(2\ell+1)}=}
            -\f{4\pi}{k}\e^{j\hj} \lt(  \phi_j^{(2\ell+1)}\bar\phi^\hi_{(2\ell+2)}\phi_\hj^{(2\ell+2)}
                                         -\phi_\hj^{(2\ell)}\bar\phi^\hi_{(2\ell)}\phi_j^{(2\ell+1)} \rt),   \\
&& \d\bar\psi_i^{(2\ell)}=-\bar\e_{i\hi}\g^\m D_\m\bar\phi^\hi_{(2\ell)} + \bar\vth_{i\hi}\bar\phi^\hi_{(2\ell)}
+\f{2\pi}{k}\bar\e_{i\hi} \lt(  \bar\phi^\hi_{(2\ell)}\phi_j^{(2\ell+1)}\bar\phi^j_{(2\ell+1)}   \rt.            \nn\\
&&\phantom{\d\psi_i^{(2\ell)}=}    \lt.  +\bar\phi^\hi_{(2\ell)}\phi_\hj^{(2\ell)}\bar\phi^\hj_{(2\ell)}
                                         -\bar\phi^j_{(2\ell-1)}\phi_j^{(2\ell-1)}\bar\phi^\hi_{(2\ell)}
                                         -\bar\phi^\hj_{(2\ell)}\phi_\hj^{(2\ell)}\bar\phi^\hi_{(2\ell)} \rt)      \nn\\
&& \phantom{\d\psi_i^{(2\ell)}=}
            +\f{4\pi}{k}\bar\e_{j\hj} \lt(  \bar\phi^j_{(2\ell-1)}\phi_i^{(2\ell-1)}\bar\phi^\hj_{(2\ell)}
                                            -\bar\phi^\hj_{(2\ell)}\phi_i^{(2\ell+1)}\bar\phi^j_{(2\ell+1)} \rt),  \nn\\
&& \d\bar\psi_\hi^{(2\ell+1)}= \bar\e_{i\hi}\g^\m D_\m\bar\phi^i_{(2\ell+1)} - \bar\vth_{i\hi}\bar\phi^i_{(2\ell+1)}
         -\f{2\pi}{k}\bar\e_{i\hi} \lt(  \bar\phi^i_{(2\ell+1)}\phi_j^{(2\ell+1)}\bar\phi^j_{(2\ell+1)}  \rt.  \nn\\
&& \phantom{\d\bar\psi_\hi^{(2\ell+1)}=} \lt. +\bar\phi^i_{(2\ell+1)}\phi_\hj^{(2\ell)}\bar\phi^\hj_{(2\ell)}
                                              -\bar\phi^j_{(2\ell+1)}\phi_j^{(2\ell+1)}\bar\phi^i_{(2\ell+1)}
                                              -\bar\phi^\hj_{(2\ell+2)}\phi_\hj^{(2\ell+2)}\bar\phi^i_{(2\ell+1)} \rt) \nn\\
&& \phantom{\d\bar\psi_\hi^{(2\ell+1)}=}
            +\f{4\pi}{k}\bar\e_{j\hj} \lt(  \bar\phi^j_{(2\ell+1)}\phi_\hi^{(2\ell)}\bar\phi^\hj_{(2\ell)}
                                            -\bar\phi^\hj_{(2\ell+2)}\phi_\hi^{(2\ell+2)}\bar\phi^j_{(2\ell+1)} \rt). \nn
\eea
Here $\ell=0,1, \cdots, r-1 $. There are no summations of $\ell$ here, and would not be summations of $\ell$ later unless it is pointed out explicitly.
Here $i,j,\cdots=1,2$ and $\hi,\hj,\cdots=\hat 1,\hat 2$ are indices of the $SU(2)\times SU(2)$ R-symmetry.
We have definitions of covariant derivatives
\bea
&& D_\m \phi_\hi^{(2\ell)} =\p_\m \phi_\hi^{(2\ell)} +\ii A_\m^{(2\ell+1)} \phi_\hi^{(2\ell)}
                                                     -\ii \phi_\hi^{(2\ell)} \hat A_\m ^{(2\ell)} ,\nn\\
&& D_\m \phi_i^{(2\ell+1)} =\p_\m \phi_i^{(2\ell+1)} +\ii A_\m^{(2\ell+1)} \phi_i^{(2\ell+1)}
                                                     -\ii \phi_i^{(2\ell+1)} \hat A_\m ^{(2\ell+2)} ,\nn\\
&& D_\m \bar\phi^\hi_{(2\ell)} =\p_\m \bar\phi^\hi_{(2\ell)} +\ii \hat A_\m ^{(2\ell)} \bar\phi^\hi_{(2\ell)}
                                                             -\ii \bar\phi^\hi_{(2\ell)}A_\m^{(2\ell+1)} ,\\
&& D_\m \bar\phi^i_{(2\ell+1)} =\p_\m \bar\phi^i_{(2\ell+1)} +\ii \hat A_\m ^{(2\ell+2)} \bar\phi^i_{(2\ell+1)}
                                                             -\ii \bar\phi^i_{(2\ell+1)}A_\m^{(2\ell+1)}. \nn
\eea
We have SUSY parameters $\e^{i\hi}=\th^{i\hi}+x^\m\g_\m\vth^{i\hi}$ and $\bar\e_{i\hi}=\bar\th_{i\hi}-\bar\vth_{i\hi}x^\m\g_\m$ with constraints
\be
(\th^{i\hi})^*=\bar \th_{i\hi}, ~~~ \bar\th_{i\hi}=\e_{ij}\e_{\hi\hj}\th^{j\hj}, ~~~
(\vth^{i\hi})^*=\bar \vth_{i\hi}, ~~~ \bar\vth_{i\hi}=\e_{ij}\e_{\hi\hj}\vth^{j\hj}.
\ee
Symbols $\e_{ij}$ and $\e_{\hi\hj}$ are antisymmetric with $\e_{12}=\e_{\hat 1 \hat 2}=1$.
The supercharges are defined as
\be \label{scne4}
\d=\ii (\bar \th_{i\hi} P^{i\hi} + \bar \vth_{i\hi} S^{i\hi})
  =\ii (\bar P_{i\hi} \th^{i\hi} + \bar S_{i\hi} \vth^{i\hi}),
\ee
with the constraints
\be
(P^{i\hi})^*=\bar P_{i\hi}, ~~~ \bar P_{i\hi}=\e_{ij}\e_{\hi\hj} P^{j\hj}, ~~~
(S^{i\hi})^*=\bar S_{i\hi}, ~~~ \bar S_{i\hi}=\e_{ij}\e_{\hi\hj} S^{j\hj}.
\ee

In Euclidean space, the SUSY transformations (\ref{susytransfne4}) and the definitions of supercharges (\ref{scne4}) still apply, but the constraints become
\be
\bar\th_{i\hi}=\e_{ij}\e_{\hi\hj}\th^{j\hj}, ~~~
\bar\vth_{i\hi}=\e_{ij}\e_{\hi\hj}\vth^{j\hj}, ~~~
\bar P_{i\hi}=\e_{ij}\e_{\hi\hj} P^{j\hj}, ~~~
\bar S_{i\hi}=\e_{ij}\e_{\hi\hj} S^{j\hj}.
\ee

\subsection{Straight line in Minkowski spacetime}

We consider BPS GY and DT type Wilson loops along timelike infinite straight lines in Minkowski spacetime.

\subsubsection{GY type Wilson loop}

There is GY type Wilson loop along the line $x^\m=\t\d^\m_0$
\bea\label{ne4gy}
&& W_\GY^{(\ell)}=\mP \exp \lt( -\ii\int\dd\t L^{(\ell)}_\GY(\t) \rt), ~~~
   L_\GY^{(\ell)}=\lt( \ba{cc} \mA_\GY^{(2\ell+1)} & \\ & \hat\mA_\GY^{(2\ell)} \ea \rt),                                   \nn\\
&& \mA_\GY^{(2\ell+1)} = A_\m^{(2\ell+1)} \dot x^\m
                       +\f{2\pi}{k} ( R^{\ph{(\ell)}i}_{(\ell)\ph ij} \phi_i^{(2\ell+1)}\bar\phi^j_{(2\ell+1)}
                                    + R^{\ph{(\ell)}\hi}_{(\ell)\ph \hi\hj} \phi_\hi^{(2\ell)}\bar\phi^\hj_{(2\ell)} )|\dot x|,\\
&& \hat\mA_\GY^{(2\ell)} = \hat A_\m^{(2\ell)} \dot x^\m
                +\f{2\pi}{k} ( S_{\ph{(\ell)}i}^{(\ell)\ph ij} \bar\phi^i_{(2\ell-1)}\phi_j^{(2\ell-1)}
                             + S_{\ph{(\ell)}\hi}^{(\ell)\ph \hi\hj} \bar\phi^\hi_{(2\ell)}\phi_\hj^{(2\ell)} )|\dot x|.\nn
\eea
The Poincar\'e SUSY transformation $\d \mA^{(2\ell+1)}_\GY=0$ leads to
\bea
&& \g_0\th^{i\hi}=-\ii R^{\ph{(\ell)}i}_{(\ell)\ph ij} \th^{j\hi}, ~~~
   \g_0\th^{i\hi}=-\ii R^{\ph{(\ell)}\hi}_{(\ell)\ph \hi\hj} \th^{i\hj},  \nn\\
&& \bar\th_{i\hi}\g_0=-\ii R^{\ph{(\ell)}j}_{(\ell)\ph ji} \bar\th_{j\hi}, ~~~
   \bar\th_{i\hi}\g_0=-\ii R^{\ph{(\ell)}\hj}_{(\ell)\ph \hj\hi} \bar\th_{i\hj}.
\eea
Taking complex conjugates of the last two equations we have
\be
\g_0\th^{i\hi}=-\ii R^{\dagger\ph{(\ell)}\hspace{-1mm} i}_{(\ell)\ph ij} \th^{j\hi}, ~~~
\g_0\th^{i\hi}=-\ii R^{\dagger\ph{(\ell)}\hspace{-1mm} \hi}_{(\ell)\ph \hi\hj} \th^{i\hj},
\ee
with $R^{\dagger\ph{(\ell)}\hspace{-1mm} i}_{(\ell)\ph ij}=(R^{\ph{(\ell)}j}_{(\ell)\ph ji})^*$ and $R^{\dagger\ph{(\ell)}\hspace{-1mm} \hi}_{(\ell)\ph \hi\hj}=(R^{\ph{(\ell)}\hj}_{(\ell)\ph \hj\hi})^*$. We define
\bea
&& B^{\ph{(\ell)}i}_{(\ell)\ph ij}= \f{R^{\ph{(\ell)}i}_{(\ell)\ph ij}+R^{\dagger\ph{(\ell)}\hspace{-1mm} i}_{(\ell)\ph ij}}{2}, ~~~
   C^{\ph{(\ell)}i}_{(\ell)\ph ij}= -\ii\f{R^{\ph{(\ell)}i}_{(\ell)\ph ij}-R^{\dagger\ph{(\ell)}\hspace{-1mm} i}_{(\ell)\ph ij}}{2},  \nn\\
&& B^{\ph{(\ell)}\hi}_{(\ell)\ph \hi\hj}
  =\f{R^{\ph{(\ell)}\hi}_{(\ell)\ph \hi\hj}+R^{\dagger\ph{(\ell)}\hspace{-1mm} \hi}_{(\ell)\ph \hi\hj}}{2}, ~~~
   C^{\ph{(\ell)}\hi}_{(\ell)\ph \hi\hj}
  = -\ii\f{R^{\ph{(\ell)}\hi}_{(\ell)\ph \hi\hj}-R^{\dagger\ph{(\ell)}\hspace{-1mm} \hi}_{(\ell)\ph \hi\hj}}{2},
\eea
and then we get
\bea
&& \g_0\th^{i\hi}=-\ii B^{\ph{(\ell)}i}_{(\ell)\ph ij} \th^{j\hi}, ~~~
   C^{\ph{(\ell)}i}_{(\ell)\ph ij} \th^{j\hi}=0,                                \nn\\
&& \g_0\th^{i\hi}=-\ii B^{\ph{(\ell)}\hi}_{(\ell)\ph \hi\hj} \th^{i\hj}, ~~~
   C^{\ph{(\ell)}\hi}_{(\ell)\ph \hi\hj} \th^{i\hj}=0.
\eea
We use $SU(2)\times SU(2)$ R-symmetry rotation and make
\be
B^{\ph{(\ell)}i}_{(\ell)\ph ij}=\diag(b_1^{(\ell)},b_2^{(\ell)}), ~~~
B^{\ph{(\ell)}\hi}_{(\ell)\ph \hi\hj}=\diag(b_{\hat 1}^{(\ell)},b_{\hat 2}^{(\ell)}).
\ee
Then we have
\be
\g_0\th^{i\hi}=-\ii b_i^{(\ell)} \th^{i\hi}, ~~~
\g_0\th^{i\hi}=-\ii b_\hi^{(\ell)} \th^{i\hi},
\ee
with no summation of indices on the right-hand sides of the two equations. Without loss of generality we choose
\be
\g_0 \th^{1\hat1}=\ii\th^{1\hat1}, ~~~\th^{1\hat1} \neq 0.
\ee
Using $\th^{1\hat1*}=\th^{2\hat2}$, we get
\be
\g_0 \th^{2\hat2}=-\ii\th^{2\hat2}, ~~~ \th^{2\hat2} \neq 0.
\ee
This leads to that $b_1^{(\ell)}=-b_2^{(\ell)}=-1$ and $b_{\hat1}^{(\ell)}=-b_{\hat2}^{(\ell)}=-1$. Furthermore, we have $\th^{1\hat2}=\th^{2\hat1}=0$. Then we get $C^{\ph{(\ell)}i}_{(\ell)\ph ij}= C^{\ph{(\ell)}\hi}_{(\ell)\ph \hi\hj}=0$.
The SUSY transformation $\d\hat\mA^{(2\ell)}_\GY=0$ leads to
\bea
&& \g_0\th^{i\hi}=-\ii S_{\ph{(\ell)}j}^{(\ell)\ph ji} \th^{j\hi}, ~~~
   \g_0\th^{i\hi}=-\ii S_{\ph{(\ell)}\hj}^{(\ell)\ph \hj\hi} \th^{i\hj},  \nn\\
&& \bar\th_{i\hi}\g_0=-\ii S_{\ph{(\ell)}i}^{(\ell)\ph ij} \bar\th_{j\hi}, ~~~
   \bar\th_{i\hi}\g_0=-\ii S_{\ph{(\ell)}\hi}^{(\ell)\ph \hi\hj} \bar\th_{i\hj}.
\eea
Then we get
\bea
&& ( R^{\ph{(\ell)}i}_{(\ell)\ph ij} - S_{\ph{(\ell)}j}^{(\ell)\ph ji} ) \th^{j\hi}=0,  ~~~
   ( R^{\ph{(\ell)}\hi}_{(\ell)\ph \hi\hj} - S_{\ph{(\ell)}\hj}^{(\ell)\ph \hj\hi} ) \th^{i\hj}=0,  \nn\\
&& ( R^{\ph{(\ell)}j}_{(\ell)\ph ji} - S_{\ph{(\ell)}i}^{(\ell)\ph ij} ) \bar\th_{j\hi}=0,  ~~~
   ( R^{\ph{(\ell)}\hj}_{(\ell)\ph \hj\hi} - S_{\ph{(\ell)}\hi}^{(\ell)\ph \hi\hj} ) \bar\th_{i\hj}=0,
\eea
and we get $S_{\ph{(\ell)}j}^{(\ell)\ph ji}=R^{\ph{(\ell)}i}_{(\ell)\ph ij}$ and $S_{\ph{(\ell)}\hj}^{(\ell)\ph \hj\hi}=R^{\ph{(\ell)}\hi}_{(\ell)\ph \hi\hj}$.

In summary, we have the GY type 1/4 BPS Wilson loop (\ref{ne4gy}) with $S_{\ph{(\ell)}j}^{(\ell)\ph ji}=R^{\ph{(\ell)}i}_{(\ell)\ph ij}=\diag(-1,1)$ and $S_{\ph{(\ell)}\hj}^{(\ell)\ph \hj\hi}=R^{\ph{(\ell)}\hi}_{(\ell)\ph \hi\hj}=\diag(-1,1)$, and the preserved supersymmetries are
\be\label{susy3}
\g_0\th^{1\hat1}=\ii\th^{1\hat1}, ~~~
\g_0\th^{2\hat2}=-\ii\th^{2\hat2}, ~~~
\th^{1\hat2}=\th^{2\hat1}=0.
\ee
This is just the 1/4 BPS Wilson loop that was constructed in \cite{Ouyang:2015qma,Cooke:2015ila}.

\subsubsection{DT type Wilson loop}

We want a DT type Wilson loop that preserves the same supersymmetries as these of the GY type 1/4 BPS Wilson loop (\ref{susy3}). A general DT type Wilson loop is
\bea
&& W_\DT^{(\ell)}=\mP \exp \lt( -\ii\int\dd\t L^{(\ell)}_\DT(\t) \rt), ~~~
   L_\DT^{(\ell)}=\lt( \ba{cc} \mA^{(2\ell+1)} & \bar f_1^{(\ell)}\\ f_2^{(\ell)} & \hat\mA^{(2\ell)} \ea \rt),     \nn\\
&& \mA^{(2\ell+1)} = \mA^{(2\ell+1)}_\GY
                       +\f{2\pi}{k} ( M^{\ph{(\ell)}i}_{(\ell)\ph ij} \phi_i^{(2\ell+1)}\bar\phi^j_{(2\ell+1)}
                                    + M^{\ph{(\ell)}\hi}_{(\ell)\ph \hi\hj}
                                      \phi_\hi^{(2\ell)}\bar\phi^\hj_{(2\ell)} )|\dot x|, \nn\\
&& \hat\mA^{(2\ell)} = \hat\mA^{(2\ell)}_\GY
                +\f{2\pi}{k} ( N_{\ph{(\ell)}i}^{(\ell)\ph ij} \bar\phi^i_{(2\ell-1)}\phi_j^{(2\ell-1)}
                              +N_{\ph{(\ell)}\hi}^{(\ell)\ph \hi\hj} \bar\phi^\hi_{(2\ell)}\phi_\hj^{(2\ell)} )|\dot x|,  \\
&& \bar f_1^{(\ell)}=\sr{\f{2\pi}{k}}\bar\zeta_i^{(\ell)}\psi^i_{(2\ell)} |\dot x|, ~~~
   f_2^{(\ell)}=\sr{\f{2\pi}{k}}\bar\psi_i^{(2\ell)}\eta^i_{(\ell)} |\dot x|.                                   \nn
\eea

From SUSY variations of $\mA^{(2\ell+1)}$ and $\hat\mA^{(2\ell)}$, we get
\bea \label{e6}
&& M^{\ph{(\ell)}i}_{(\ell)\ph ij} =N_{\ph{(\ell)}j}^{(\ell)\ph ji}=0, ~~~
   M^{\ph{(\ell)}\hi}_{(\ell)\ph \hi\hj} =N_{\ph{(\ell)}\hj}^{(\ell)\ph \hj\hi},\\
&& \sr{\f{2\pi}{k}} M^{\ph{(\ell)}\hi}_{(\ell)\ph \hi\hj} \phi^{(2\ell)}_\hi \th^{i\hj}=-\eta^i_{(\ell)} \bar g_1^{(\ell)}, ~~~
   \sr{\f{2\pi}{k}} M^{\ph{(\ell)}\hi}_{(\ell)\ph \hi\hj} \bar\phi^\hj_{(2\ell)} \bar \th_{i\hi}=g_2^{(\ell)}\bar\z^{(\ell)}_i.\nn
\eea
From variations of $\bar f_1^{(\ell)}$ and $f_2^{(\ell)}$ we get
\bea\label{fermiorbifold}
&& \bar\zeta^{(\ell)}_1\g_0=\ii\bar\zeta^{(\ell)}_1, ~~~ \bar\zeta^{(\ell)}_2\g_0=-\ii\bar\zeta^{(\ell)}_2, ~~~
   \g_0 \eta_{(\ell)}^1=\ii \eta_{(\ell)}^1, ~~~ \g_0 \eta_{(\ell)}^2=-\ii \eta_{(\ell)}^2,                       \nn\\
&& \bar g^{(\ell)}_1 = -\sr{\f{2\pi}{k}} \bar\zeta^{(\ell)}_i \g_0\th^{i\hi}\phi^{(2\ell)}_\hi,  ~~~
   M^{\ph{(\ell)}\hi}_{(\ell)\ph\hi\hj} \bar\z_i^{(\ell)} \gamma_0\th^{i\hk}
  =M^{\ph{(\ell)}\hk}_{(\ell)\ph\hk\hj} \bar\z_i^{(\ell)} \gamma_0\theta^{i\hi},                       \\
&& g^{(\ell)}_2 = \sr{\f{2\pi}{k}} \bar\th_{i\hi}\g_0\eta_{(\ell)}^i\bar\phi_{(2\ell)}^\hi,                                  ~~~
   M^{\ph{(\ell)}\hi}_{(\ell)\ph \hi\hj} \bar\th_{i\hk}\gamma_0\eta^i_{(\ell)}
  =M^{\ph{(\ell)}\hi}_{(\ell)\ph \hi\hk} \bar\th_{i\hj}\gamma_0\eta^i_{(\ell)}.                                   \nn
\eea
We make the parameterizations
\bea
&& \bar\z_1^{(\ell)}=\bar\a^{(\ell)}\bar\z, ~~~ \bar \z^\a=(1,\ii),    ~~~
   \bar\z_2^{(\ell)}=\bar\g^{(\ell)}\bar\m, ~~~ \bar \m^\a=(-\ii,-1),   \nn\\
&& \eta^1_{(\ell)}=\eta\b_{(\ell)}, ~~~ \eta_\a=(1,-\ii),               ~~~
  \eta^2 _{(\ell)}=\n\d_{(\ell)}, ~~~ \n_\a=(-\ii,1).
\eea
Equations (\ref{e6}) become
\be
M^{\ph{(\ell)}\hi}_{(\ell)\ph \hi\hj} \th^{i\hj}=\eta_{(\ell)}^i\bar\zeta^{(\ell)}_j\g_0\th^{j\hi}, ~~~
M^{\ph{(\ell)}\hj}_{(\ell)\ph \hj\hi} \bar\th_{i\hj}= \bar\th_{j\hi} \g_0 \eta_{(\ell)}^j \bar\zeta^{(\ell)}_i,
\ee
from which we have
\bea
&& M^{\ph{(\ell)}\hat1}_{(\ell)\ph{\hat1}\hat1} = 2\ii\bar\a^{(\ell)}\b_{(\ell)},   ~~~
   M^{\ph{(\ell)}\hat2}_{(\ell)\ph{\hat2}\hat2} = 2\ii\bar\g^{(\ell)}\d_{(\ell)},  \nn\\
&& M^{\ph{(\ell)}\hat1}_{(\ell)\ph{\hat1}\hat2} =
   M^{\ph{(\ell)}\hat2}_{(\ell)\ph{\hat2}\hat1} = \bar\a^{(\ell)}\d_{(\ell)} = \bar\g^{(\ell)}\b_{(\ell)}=0.
\eea
We have four classes of solutions.

\subsubsection*{Class I}

In the first solution we have
\be
\bar\g^{(\ell)}=\d_{(\ell)}=0.
\ee
We have the DT type 1/4 BPS Wilson loop that preserves the supersymmetries (\ref{susy3}),
and there are two free complex parameters $\bar\a^{(\ell)}$ and $\b_{(\ell)}$.
Now we have
\bea
&& L_B^{(\ell)} = \f{4\pi\ii}{k} \bar\a^{(\ell)}\b_{(\ell)}
         \lt( \ba{cc}   \phi_{\hat1}^{(2\ell)}\bar\phi^{\hat1}_{(2\ell)} &  \\
                      & \bar\phi^{\hat1}_{(2\ell)}\phi_{\hat1}^{(2\ell)}   \ea \rt),~~~
   L_F^{(\ell)} = \sr{\f{2\pi}{k}} \lt( \ba{cc} & \bar\a^{(\ell)}\bar\zeta \psi^1_{(2\ell)}
                                            \\ \bar\psi_1^{(2\ell)} \eta\b_{(\ell)}  & \ea \rt), \nn\\
&& \L^{(\ell)} = -\sr{\f{2\pi}{k}} \lt( \ba{cc} & \bar\a^{(\ell)} \phi_{\hat1}^{(2\ell)}
                                             \\ \b_{(\ell)} \bar\phi^{\hat1}_{(2\ell)} & \ea \rt), ~~~
   \k=2\ii, ~~~
   Q=\bar\zeta P^{1\hat1}+\bar P_{1\hat1}\eta.
\eea
This shows that difference between the DT type Wilson loop and GY type Wilson loop is $Q$-exact.

We search for SUSY enhancement of the DT type BPS Wilson loop. One of the consequences of SUSY invariance of the Wilson loop is that
\be
\g_0\th^{i\hi}=-\ii R^{\ph{(\ell)}i}_{(\ell)\ph ij} \th^{j\hi}, ~~~
\bar\th_{i\hi}\g_0=-\ii R^{\ph{(\ell)}j}_{(\ell)\ph ji} \bar\th_{j\hi},
\ee
from which we have
\be \label{ss2}
\g_0\th^{1\hi}=\ii \th^{1\hi}, ~~~
\g_0\th^{2\hi}=-\ii\th^{2\hi}, ~~~
\hi=\hat 1,\hat2.
\ee
From \cite{Ouyang:2015hta} we know that the only possibility of SUSY enhancement in the present case is that when
\be
\bar\a^{(\ell)} \b_{(\ell)}=-\ii,
\ee
and supersymmetries are enhanced to 1/2 BPS. The preserved supersymmetries are (\ref{ss2}).
This is just the $\psi_1$-loop that was constructed in \cite{Ouyang:2015qma,Cooke:2015ila}.

\subsubsection*{Class II}

In the second solution we have
\be
\bar\a^{(\ell)}=\b_{(\ell)}=0.
\ee
We have the DT type 1/4 BPS Wilson loop that preserves the supersymmetries (\ref{susy3}).
Now we have
\bea
&& L_B^{(\ell)} = \f{4\pi\ii}{k} \bar\g^{(\ell)}\d_{(\ell)}
         \lt( \ba{cc}   \phi_{\hat2}^{(2\ell)}\bar\phi^{\hat2}_{(2\ell)} &  \\
                      & \bar\phi^{\hat2}_{(2\ell)}\phi_{\hat2}^{(2\ell)}   \ea \rt),~~~
   L_F^{(\ell)} = \sr{\f{2\pi}{k}} \lt( \ba{cc} & \bar\g^{(\ell)}\bar\m \psi^2_{(2\ell)} \\ \bar\psi_2^{(2\ell)} \n\d_{(\ell)}  & \ea \rt), \nn\\
&& \L^{(\ell)} = -\sr{\f{2\pi}{k}} \lt( \ba{cc} & \bar\g^{(\ell)} \phi_{\hat2}^{(2\ell)}
                                             \\ \d_{(\ell)} \bar\phi^{\hat2}_{(2\ell)} & \ea \rt), ~~~
   \k=2\ii, ~~~
   Q=\bar\m P^{2\hat2}+\bar P_{2\hat2}\n.
\eea
When
\be
\bar\g^{(\ell)} \d_{(\ell)}=\ii,
\ee
supersymmetries are enhanced to 1/2 BPS. The preserved supersymmetries are also (\ref{ss2}).
This is just the $\psi_2$-loop that was constructed in \cite{Cooke:2015ila}. The $\psi_1$-loop and $\psi_2$-loop have the same preserved supersymmetries.

\subsubsection*{Class III}

In the third solution we have
\be
\b_{(\ell)}=\d_{(\ell)}=0.
\ee
Now we have
\bea
&& L_B^{(\ell)} =0,~~~
   L_F^{(\ell)} = \sr{\f{2\pi}{k}} \lt( \ba{cc} & \bar\a^{(\ell)}\bar\zeta \psi^1_{(2\ell)}+\bar\g^{(\ell)}\bar\n \psi^2_{(2\ell)} \\ 0  & \ea \rt), \nn\\
&& \L^{(\ell)} = -\sr{\f{2\pi}{k}} \lt( \ba{cc} & \bar\a^{(\ell)} \phi_{\hat1}^{(2\ell)} +\bar\g^{(\ell)} \phi_{\hat2}^{(2\ell)}
                                             \\ 0 & \ea \rt),\\
&& \k=-2\ii, ~~~
   Q=\bar\zeta P^{1\hat1}+\bar\m P^{2\hat2}.     \nn
\eea
There is no SUSY enhancement in the present case.

\subsubsection*{Class IV}

In the fourth solution we have
\be
\bar\a^{(\ell)}=\bar\g^{(\ell)}=0.
\ee
Now we have
\bea
&& L_B^{(\ell)} =0,~~~
   L_F^{(\ell)} = \sr{\f{2\pi}{k}} \lt( \ba{cc} &  0  \\ \bar\psi_1^{(2\ell)} \eta\b_{(\ell)} +\bar\psi_2^{(2\ell)} \n\d_{(\ell)}  & \ea \rt), \nn\\
&& \L^{(\ell)} = -\sr{\f{2\pi}{k}}
                  \lt( \ba{cc} & 0 \\ \b_{(\ell)} \bar\phi^{\hat1}_{(2\ell)} + \d_{(\ell)} \bar\phi^{\hat2}_{(2\ell)} & \ea \rt),\\
&& \k=-2\ii, ~~~
   Q=\bar P_{1\hat1}\eta+\bar P_{2\hat2}\n.\nn
\eea
There is no SUSY enhancement in the present case.

\subsection{Circle in Euclidean space}

We consider circular BPS GY and DT type Wilson loops in Euclidean space.

\subsubsection{GY type Wilson loop}

In Euclidean space we have the 1/4 BPS circular GY type Wilson loops along $x^\m=(\cos\t,\sin\t,0)$
\bea
&& W_\GY^{(\ell)}=\Tr\mP \exp \lt( -\ii\oint\dd\t L^{(\ell)}_\GY(\t) \rt), ~~~
   L_\GY^{(\ell)}=\lt( \ba{cc} \mA_\GY^{(2\ell+1)} & \\ & \hat\mA_\GY^{(2\ell)} \ea \rt),                                   \nn\\
&& \mA_\GY^{(2\ell+1)} = A_\m^{(2\ell+1)} \dot x^\m
                       +\f{2\pi}{k} ( R^{\ph{(\ell)}i}_{(\ell)\ph ij} \phi_i^{(2\ell+1)}\bar\phi^j_{(2\ell+1)}
                                    + R^{\ph{(\ell)}\hi}_{(\ell)\ph \hi\hj} \phi_\hi^{(2\ell)}\bar\phi^\hj_{(2\ell)} )|\dot x|,\\
&& \hat\mA_\GY^{(2\ell)} = \hat A_\m^{(2\ell)} \dot x^\m
                +\f{2\pi}{k} ( R^{\ph{(\ell)}i}_{(\ell)\ph ij} \bar\phi^j_{(2\ell-1)}\phi_i^{(2\ell-1)}
                             + R^{\ph{(\ell)}\hi}_{(\ell)\ph \hi\hj} \bar\phi^\hj_{(2\ell)}\phi_\hi^{(2\ell)} )|\dot x|, \nn
\eea
with $R^{\ph{(\ell)}i}_{(\ell)\ph ij}=R^{\ph{(\ell)}\hi}_{(\ell)\ph \hi\hj}=\diag(\ii,-\ii)$. The preserved supersymmetries are
\be \label{susyne4circ}
\vth^{1\hat1}=\ii\g_3\th^{1\hat1}, ~~~ \vth^{2\hat2}=-\ii\g_3\th^{2\hat2}, ~~~ \th^{1\hat2}=\th^{2\hat1}=\vth^{1\hat2}=\vth^{2\hat1}=0.
\ee

\subsubsection{DT type Wilson loop}

Along $x^\m=(\cos\t,\sin\t,0)$ we construct the circular DT type Wilson loop
\bea
&& W_\DT^{(\ell)}=\Tr\mP \exp \lt( -\ii\oint\dd\t L^{(\ell)}_\DT(\t) \rt), ~~~
   L_\DT^{(\ell)}=\lt( \ba{cc} \mA^{(2\ell+1)} & \bar f_1^{(\ell)}\\ f_2^{(\ell)} & \hat\mA^{(2\ell)} \ea \rt),     \nn\\
&& \mA^{(2\ell+1)} = \mA^{(2\ell+1)}_\GY
                       -\f{4\pi}{k} (   \bar\a^{(\ell)}\b_{(\ell)} \phi_{\hat1}^{(2\ell)}\bar\phi^{\hat1}_{(2\ell)}
                                      + \bar\g^{(\ell)}\d_{(\ell)} \phi_{\hat2}^{(2\ell)}\bar\phi^{\hat2}_{(2\ell)} )|\dot x|, \nn\\
&& \hat\mA^{(2\ell)} = \hat\mA^{(2\ell)}_\GY
                -\f{4\pi}{k} (   \bar\a^{(\ell)}\b_{(\ell)}\bar\phi^{\hat1}_{(2\ell)}\phi_{\hat1}^{(2\ell)}
                                      + \bar\g^{(\ell)}\d_{(\ell)}\bar\phi^{\hat2}_{(2\ell)}\phi_{\hat2}^{(2\ell)} )|\dot x|,\\
&& \bar f_1^{(\ell)}=\sr{\f{2\pi}{k}} (  \bar\a^{(\ell)}\bar\zeta \psi^1_{(2\ell)}+\bar\g^{(\ell)}\bar\m \psi^2_{(2\ell)} ) |\dot x|, ~~~
   \bar\z^\a=(\ep^{\ii\t/2},\ep^{-\ii\t/2}), ~~~
   \bar\m^\a=(\ep^{\ii\t/2},-\ep^{-\ii\t/2}),  \nn\\
&& f_2^{(\ell)}=\sr{\f{2\pi}{k}} ( \bar\psi_1^{(2\ell)}\eta\b_{(\ell)} + \bar\psi_2^{(2\ell)}\n\d_{(\ell)} )|\dot x|, ~~~
   \eta_\a=(\ep^{-\ii\t/2},\ep^{\ii\t/2}),  ~~~
   \n_\a=(-\ep^{-\ii\t/2},\ep^{\ii\t/2}). \nn
\eea
To make it preserve the supersymmetries (\ref{susyne4circ}), we have four classes of solutions, and all of them satisfy
\be
\bar\a^{(\ell)}\d_{(\ell)}=\bar\g^{(\ell)}\b_{(\ell)}=0.
\ee

\subsubsection*{Class I}

In the first solution we have
\be
\bar\g^{(\ell)}=\d_{(\ell)}=0.
\ee
Now we have
\bea
&& L_B^{(\ell)} = -\f{4\pi}{k} \bar\a^{(\ell)}\b_{(\ell)}
         \lt( \ba{cc}   \phi_{\hat1}^{(2\ell)}\bar\phi^{\hat1}_{(2\ell)} &  \\
                      & \bar\phi^{\hat1}_{(2\ell)}\phi_{\hat1}^{(2\ell)}   \ea \rt),~~~
   L_F^{(\ell)} = \sr{\f{2\pi}{k}} \lt( \ba{cc} & \bar\a^{(\ell)}\bar\zeta \psi^1_{(2\ell)}
                                            \\ \bar\psi_1^{(2\ell)} \eta\b_{(\ell)}  & \ea \rt),            \nn\\
&& \L^{(\ell)} = -\sr{\f{2\pi}{k}} \ep^{\ii\t/2} \lt( \ba{cc} & \bar\a^{(\ell)} \phi_{\hat1}^{(2\ell)}
                                             \\ \b_{(\ell)} \bar\phi^{\hat1}_{(2\ell)} & \ea \rt), ~~~
   \k=-2\ep^{-\ii\t},                                                                                       \\
&& Q= \bar a (P^{1\hat1} +\ii\g_3 S^{1\hat1} ) + ( \bar P_{1\hat1} + \bar S_{1\hat1} \ii\g_3)b, ~~~
   \bar a^\a=(1,0), ~~~ b_\a=(0,1).                                                                         \nn
\eea
When
\be
\bar\a^{(\ell)} \b_{(\ell)}=\ii,
\ee
supersymmetries are enhanced to 1/2 BPS. The preserved supersymmetries are
\be \label{susy5}
\vth^{1\hi}=\ii\g_3\th^{1\hi}, ~~~
\vth^{2\hi}=-\ii\g_3\th^{2\hi}, ~~~
\hi=\hat1,\hat2.
\ee
This is just the circular $\psi_1$-loop.

\subsubsection*{Class II}

In the second solution we have
\be
\bar\a^{(\ell)}=\b_{(\ell)}=0.
\ee
Now we have
\bea
&& L_B^{(\ell)} = -\f{4\pi}{k} \bar\g^{(\ell)}\d_{(\ell)}
         \lt( \ba{cc}   \phi_{\hat2}^{(2\ell)}\bar\phi^{\hat2}_{(2\ell)} &  \\
                      & \bar\phi^{\hat2}_{(2\ell)}\phi_{\hat2}^{(2\ell)}   \ea \rt),~~~
   L_F^{(\ell)} = \sr{\f{2\pi}{k}} \lt( \ba{cc} & \bar\g^{(\ell)}\bar\m \psi^2_{(2\ell)} \\ \bar\psi_2^{(2\ell)} \n\d_{(\ell)}  & \ea \rt), \nn\\
&& \L^{(\ell)} = -\sr{\f{2\pi}{k}} \ep^{\ii\t/2}\lt( \ba{cc} & \bar\g^{(\ell)} \phi_{\hat2}^{(2\ell)}
                                             \\ \d_{(\ell)} \bar\phi^{\hat2}_{(2\ell)} & \ea \rt), ~~~
   \k=-2\ep^{-\ii\t},                                                                                      \\
&& Q= \bar a (P^{2\hat2} - \ii\g_3 S^{2\hat2} ) + ( \bar P_{2\hat2} - \bar S_{2\hat2} \ii\g_3)b, ~~~
   \bar a^\a=(1,0), ~~~ b_\a=(0,1).                                                                         \nn
\eea
When
\be
\bar\g^{(\ell)} \d_{(\ell)}=-\ii,
\ee
supersymmetries are enhanced to 1/2 BPS. The preserved supersymmetries are also (\ref{susy5}).
This is just the circular $\psi_2$-loop.

\subsubsection*{Class III}

In the third solution we have
\be
\b_{(\ell)}=\d_{(\ell)}=0.
\ee
Now we have
\bea
&& L_B^{(\ell)} =0,~~~
   L_F^{(\ell)} = \sr{\f{2\pi}{k}} \lt( \ba{cc} & \bar\a^{(\ell)}\bar\zeta \psi^1_{(2\ell)}+\bar\g^{(\ell)}\bar\n \psi^2_{(2\ell)} \\ 0  & \ea \rt), \nn\\
&& \L^{(\ell)} = -\sr{\f{2\pi}{k}} \ep^{\ii\t/2}\lt( \ba{cc} & \bar\a^{(\ell)} \phi_{\hat1}^{(2\ell)} +\bar\g^{(\ell)} \phi_{\hat2}^{(2\ell)}
                                             \\ 0 & \ea \rt), ~~~
  \k=2\ep^{-\ii\t},                                                                                      \\
&& Q= \bar a (P^{1\hat1} +\ii\g_3 S^{1\hat1} + P^{2\hat2} - \ii\g_3 S^{2\hat2} ) , ~~~
   \bar a^\a=(1,0).                                                                         \nn
\eea

\subsubsection*{Class IV}

In the fourth solution we have
\be
\bar\a^{(\ell)}=\bar\g^{(\ell)}=0.
\ee
Now we have
\bea
&& L_B^{(\ell)} =0,~~~
   L_F^{(\ell)} = \sr{\f{2\pi}{k}} \lt( \ba{cc} &  0  \\ \bar\psi_1^{(2\ell)} \eta\b_{(\ell)} +\bar\psi_2^{(2\ell)} \n\d_{(\ell)}  & \ea \rt), \nn\\
&& \L^{(\ell)} = -\sr{\f{2\pi}{k}}\ep^{\ii\t/2}
                  \lt( \ba{cc} & 0 \\ \b_{(\ell)} \bar\phi^{\hat1}_{(2\ell)} + \d_{(\ell)} \bar\phi^{\hat2}_{(2\ell)} & \ea \rt), ~~~
  \k=2\ep^{-\ii\t},                                                                                       \\
&& Q=( \bar P_{1\hat1} + \bar S_{1\hat1}\ii\g_3 + \bar P_{2\hat2} - \bar S_{2\hat2} \ii\g_3)b, ~~~
   b_\a=(0,1).                                                                         \nn
\eea

\section{Conclusion and discussion}\label{s7}

In this paper, we have constructed novel BPS Wilson loops along infinite straight lines and circles in several three-dimensional quiver superconformal CSM theories, especially the novel DT type BPS Wilson loops in ABJM theory and $\mN=4$ orbifold ABJM theory.
There are several free complex parameters in these Wilson loops.
There are SUSY enhancements at special values of the parameters for the DT type BPS Wilson loops in ABJM theory and $\mN=4$ orbifold ABJM theory.
The construction here can be generalized for DT type BPS Wilson loops along circles.
We  also notice that our construction of DT type BPS Wilson loops in  $\mN=4$ orbifold ABJM theory should be easily generalized to the one of  similar Wilson loops in  $\mN=4$ necklace quiver theory with gauge group $\prod_{i=1}^{2r} U(N_i)$ and Chern-Simons levels $(k, -k, \cdots, k, -k)$.

For the 1/6 BPS DT type Wilson loops in ABJM theory and 1/4 BPS DT type Wilson loops in $\mN=4$ orbifold ABJM theory, there are infinite degeneracies. We have an infinite number of Wilson loops along the same curve that preserve the same supersymmetries. In the spirit of \cite{Cooke:2015ila}, it is possible that not all of these Wilson loops are BPS quantum mechanically.\footnote{Recently it was found \cite{Griguolo:2015swa} that such degeneracies, at the level of vacuum expectation values, for the previously-constructed half BPS Wilson loops, the $\psi_1$- and $\psi_2$-loops, in $\mN=4$ orbifold ABJM theory are not lifted at two-loop level in the perturbation expansion. Possibility of lift at three-loop level was also discussed there.} If it is the case one should find how the degeneracies are lifted. If it is not the case, one should identify their gravity duals.

\textbf{Note added in proofreading.} Recently in \cite{Bianchi:2016vvm}, it was found that degeneracy of vacuum expectation values of half-BPS $\psi_1$- and $\psi_2$-loops, mentioned in the footnote of this section, is lifted at three loops. And strong evidences were given that the average of these two Wilson loops are half-BPS quantum mechanically.

\section*{Acknowledgement}

We would like to thank Nan Bai, Bin Chen, Nadav Drukker, Song He, Yu Jia, Wei Li, Jian-Xin Lu, Gary Shiu, Zhao-Long Wang and Piljin Yi for valuable discussions.
JW would like to thank IMP, Northwest University for warm hospitality.
The work was in part supported by National Natural Science Foundation of China Grants No.~11222549 and No.~11575202. JW also gratefully acknowledges the support of K.~C.~Wong Education Foundation.
JW would also like to thank the participants of the advanced workshop ``Dark Energy and Fundamental Theory'' supported by the Special Fund for Theoretical Physics from the National Natural Science Foundation of China with Grant No.~11447613 for stimulating discussion.


\providecommand{\href}[2]{#2}\begingroup\raggedright\endgroup


\end{document}